\documentclass[12pt]{article}
\usepackage{amsmath,  amsthm}
\usepackage{graphicx}
\usepackage{url}  
\usepackage{amssymb} 
\usepackage{color}
\usepackage{tikz}
\usepackage{caption, subcaption}
\usetikzlibrary{shapes.misc, positioning}
\newsavebox{\tempbox}
\usepackage[utf8]{inputenc}
\usepackage{graphicx}
\usepackage{natbib}
\usepackage{float}
\usepackage{caption}
\usepackage{subcaption}
\usepackage{amsmath}
\usepackage{graphicx} 
\usepackage{booktabs}
\usepackage{apalike}
\usepackage{amsmath,amsfonts,amssymb,amsthm}
\usepackage{xcolor}
\usepackage{titling}
\usepackage{lipsum}
\usepackage{multirow}
\usepackage{tabu}
\usepackage{mathtools}
\usepackage[thinlines]{easytable}
\usepackage{array}
\usepackage[title]{appendix}
\usepackage{afterpage} 

\usepackage{algorithm}
\usepackage{algorithmic}
\usepackage{tikz}
\usepackage{bm}
\usepackage{color, colortbl}
\usepackage{hhline}
\usetikzlibrary{arrows,matrix,positioning,fit, decorations.pathreplacing}
\usepackage{titlesec}
\titleformat*{\section}{\large\bfseries}
\titleformat*{\subsection}{\normalsize\bfseries}
\titleformat*{\subsubsection}{\normalsize\bfseries}
\titleformat*{\paragraph}{\normalsize\bfseries}
\titleformat*{\subparagraph}{\normalsize\bfseries}
\usepackage{setspace}
\usepackage{longtable}
\usepackage{threeparttable}

\let\proglang=\textsf

\theoremstyle{definition}
\newtheorem{remark}{Remark}[section] 
\usepackage{url}

\usepackage[normalem]{ulem}
\usepackage{color}
\newcommand{\revision}[1]{{\textcolor{black}{#1}}}

\newcommand{\textbox}[1]
{\savebox{\tempbox}{#1}
	\ifdim\wd\tempbox<4cm\relax
	\makebox[4cm]{\usebox{\tempbox}}%
	\else
	\parbox{4cm}{\raggedright #1}%
	\fi}
\newcommand{\blind}{0}

\begin{document}
	\def\spacingset#1{\renewcommand{\baselinestretch}%
		{#1}\small\normalsize} \spacingset{1}

	\if0\blind
	{
		\title{\bf Hierarchical Regularizers for \\Mixed-Frequency Vector Autoregressions}
		\author{Alain Hecq\hspace{.2cm}\\
			Department of Quantitative Economics,  Maastricht University\\
			Marie Ternes\thanks{
				Corresponding author: Marie Ternes, Maastricht University, School of Business and Economics, P.O. Box 616, 6200 MD Maastricht, The Netherlands, Email: m.ternes@maastrichtuniversity.nl} \\
		Department of Quantitative Economics,  Maastricht University
			and \\
			Ines Wilms
			\\
		Department of Quantitative Economics,  Maastricht University}
		\maketitle
	} \fi
	
	\if1\blind
	{
		\bigskip
		\bigskip
		\bigskip
		\begin{center}
			{\LARGE\bf Hierarchical Regularizers for Mixed-Frequency Vector Autoregressions}
		\end{center}
		\medskip
	} \fi
	
	\bigskip
	\begin{abstract}
Mixed-frequency Vector AutoRegressions (MF-VAR) model the dynamics between variables recorded at different frequencies. However, as the number of series and high-frequency observations per low-frequency period grow, MF-VARs suffer from the ``curse of dimensionality". We curb this curse through a regularizer that permits hierarchical sparsity patterns by prioritizing the inclusion of coefficients according to the recency of the information they contain. Additionally, we investigate the presence of nowcasting relations by sparsely estimating the MF-VAR error covariance matrix. We study predictive Granger causality relations in a MF-VAR 
for the U.S.\ economy and construct a coincident indicator of GDP growth. \revision{Supplementary Materials for this article are available online.}
	\end{abstract}
	
	\noindent%
	\vfill
	
	\spacingset{1.5} 

\section{Introduction}
Vector AutoRegressive (VAR) models are a cornerstone for modeling multivariate time series;  studying their dynamics
and  forecasting. 
However, standard VARs require all component series to enter the model at the same frequency, while in practice macro and financial series are typically recorded at different frequencies; quarterly, monthly, or weekly for instance.
One could aggregate high-frequency variables to one common low frequency and continue the analysis with a standard VAR, but such a practice wastes valuable information contained in high-frequency data \revision{due to  two main reasons. First, high-frequency data are inherently more timely; they closely track the state of the economy in real time. Second, 
they can help unmask dynamics that would be hidden under temporal aggregation (see e.g., recent discussions in \citealp{cimadomo2021nowcasting, paccagnini2021identifying}).
}

Mixed-frequency (MF) models, instead, exploit the information available in series recorded at different frequencies. One commonly used MF models is the MIxed DAta Sampling (MIDAS) regression (\citealp{ghysels2004midas}). While the literature first focused on a single-equation framework for modelling the low-frequency response, the multivariate extension by \cite{ghysels2016macroeconomics} enabled one to model the relations between high- and low-frequency series in a mixed-frequency VAR (MF-VAR) system. \revision{This is known as the stacked MF-VAR approach since the MF-VAR is estimated at the lowest frequency and all higher-frequency variables are treated as separate components series which are stacked in the MF-VAR.}\footnote{\revision{Alternatively to stacking, MF-VARs can be accommodated within the framework of state-space models, where the low-frequency variables are modelled as high-frequency series with latent observations} (e.g., \citealp{kuzin2011midas}; \citealp{foroni2014comparison}; \citealp{schorfheide2015real}; \citealp{brave2019forecasting}; \citealp{gefang2020computationally}; \citealp{koelbl2020new}; \revision{the latent ``L-BVAR" of \citealp{cimadomo2021nowcasting}}). 
We contribute to the literature stream of \cite{ghysels2016macroeconomics} as the stacked MF-VAR system allows for the application of standard VAR tools such as Granger causality to the mixed-frequency setting, as discussed in Section \ref{Methodology}.} 

A complication with MF-VARs is that they are severely affected by the ``curse of dimensionality". This curse arises due to two  sources. First, the number of parameters grows quadratically with the number of component series, just like for standard VARs. Secondly, specific to MF-VARs, we have many high-frequency observations per low-frequency observation, which each enter as different component series in the model, thereby adding to the dimensionality. 
Without further adjustments, one would be limited to MF-VARs with few  series and/or  a small number of high-frequency observations per low-frequency observation. 

This curse of dimensionality has mostly been addressed through mixed-frequency factor models (e.g., \citealp{marcellino2010factor}; \citealp{foroni2014comparison}; \citealp{andreou2019inference}) or Bayesian estimation (e.g.,  \citealp{schorfheide2015real};  \citealp{ghysels2016macroeconomics};
\citealp{gotz2016testing}; 
\citealp{mccracken2020real};
\revision{\citealp{cimadomo2021nowcasting}; \citealp{paccagnini2021identifying}}).
Sparsity-inducing 
regularizers form an appealing alternative (see \citealp{hastie2015statistical} for an introduction), but despite their popularity in regression and standard VAR settings (e.g., \citealp{hsu2008subset, basu2015regularized, basu2015network, davis2016sparse, callot2017modeling, Demirer17, Smeekes18, barigozzi2019nets, hecq2020granger}), they have only been rarely explored as a tool for dimension reduction in mixed-frequency models. 
An exception is \cite{babii2021machine} who recently used the sparse-group lasso 
to accommodate the dynamic nature of high-dimensional, mixed-frequency data, 
\revision{thereby providing a complementary structured machine learning perspective to the penalized Bayesian MIDAS approach of \cite{mogliani2021bayesian}.}
Nonetheless, they address univariate MIDAS regressions, leaving
regularization of MF-VARs unexplored. 

\revision{Our paper's first contribution concerns the introduction of a novel convex regularizer 
that extends the univariate MIDAS approach of \cite{babii2021machine} to the multivariate MF-VAR setting.
To this end, we propose a mixed-frequency extension of
the single-frequency hierarchical regularizer by \cite{nicholson2020high} used for standard VARs}
that accounts for covariates at different (high-frequency) lags being temporally ordered. We build upon the group lasso with nested groups
and encourage a hierarchical sparsity pattern that prioritizes the inclusion of coefficients according to 
 the recency of the information the corresponding series contains about the state of the economy. 

In addition to the development of a new MF-VAR with hierarchical lag structure, our paper investigates the presence of nowcasting restrictions in a high-dimensional mixed-frequency setting. 
According to Eurostat's glossary, a nowcast is ``\textit{a rapid} [estimate] \textit{produced during the current reference period} [, say $T^{\ast }$ a particular quarter$,$] \textit{for a hard economic variable of interest observed for the same reference period} [$T^{\ast}$]" \citep{Eurostatnowcast}.
In this narrow sense, and contrarily to forecasting,
nowcasting makes use of all available information becoming available between (and strictly speaking not including) $T^{\ast}-1$ and $T^{\ast}$. 
\cite{gotz2014nowcasting} show that nowcasting in (low-dimensional) MF-VARs can be studied through contemporaneous Granger causality tests; by testing the null of a block diagonal error  covariance matrix of the MF-VAR. 
We build on \cite{gotz2014nowcasting} to study nowcasting relations in high-dimensional MF-VARs by sparsely estimating the covariance matrix of the MF-VAR errors. Its sparsity pattern then provides evidence on those high-frequency variables (i.e.,\ the series and their particular time period) one can use to build coincident indicators for the low-frequency main economic indicators.\footnote{\revision{Alternatively, nowcasts can be obtained as forecasts conditional on the real-time data flow via state-space techniques such as the Kalman filter which is needed to estimate the latent processes. We discuss the method of \cite{cimadomo2021nowcasting} as state-of-the-art benchmark in this literature in Section \ref{subsection Nowcasting GDP pre and post Covid}.}}

In a simulation study (Section \ref{Simulation Study}), we find that the hierarchical regularizer performs well in terms of estimation accuracy and variable selection when compared to alternative methods. Furthermore, we accurately retrieve nowcasting relations between the low- and high-frequency variables by sparsely estimating the error covariance matrix. 
In the application (Section \ref{Macroeconomic Application}), we study a high-dimensional MF-VAR for the U.S.\ economy. 
We apply the hierarchical regularizer to characterize the predictive Granger causality relations through a network analysis. Moreover, we investigate which high-frequency  series nowcast quarterly U.S.\ real gross domestic product (GDP) growth, use those to construct a reliable coincident indicator of GDP growth and \revision{evaluate its performance pre and post Covid-19.}

The remainder of this paper is structured as follows. 
Section \ref{Methodology} introduces the MF-VAR with hierarchically structured parameters and defines the nowcasting causality relations, Section \ref{Regularized Estimation Procedure of MF-VARs} describes the regularized estimation procedure for MF-VARs and nowcasting causality. 
Section \ref{Simulation Study} shows the results on the simulation study, Section \ref{Macroeconomic Application} on the empirical application of the U.S.\ economy. Section \ref{Conclusion} concludes. \revision{Additional results are available in the online appendix.}

\section{Mixed-Frequency VARs} \label{Methodology}
We start from \citeauthor{ghysels2016macroeconomics}' (\citeyear{ghysels2016macroeconomics}) mixed-frequency VAR systems and include $d$ different high-frequency components. Let $\mathbf{y}(t)$ denote a  $k_L$-dimensional vector collecting the low-frequency variables for $t=1, \dots, T$. Further, let $\mathbf{x}^{m_1}(t)$,  $\mathbf{x}^{m_2}(t)$,...,$\mathbf{x}^{m_d}(t)$ denote the $d$ different, multivariate high-frequency components with $m_1, \ m_2, \dots, m_d$ number of high-frequency observations per low-frequency period $t$, respectively. Without loss of generality, we take $m_1 < m_2 < \dots < m_d$. For instance, in a quarter/month/weekly-example $m_1=3$ and $m_2=12$. Let each component contain $k_1, k_2, \dots, k_d$ time series of the same high-frequency, respectively. To be precise, each high-frequency component is given by 
$
    \mathbf{x}^{m_i}(t) = [x_{1}(t,m_i), \dots, x_{1}(t,1), \dots, x_{k_i}(t,m_i), \dots, x_{k_i}(t,1)]^{\prime} \in \mathbb{R}^{m_i\cdot k_i}
$
(for $i=1,\dots,d$),
where the couple $(t,j)$ indicates higher-frequency period $j=1, \dots, m_i$ during the low-frequency period $t$.
The MF-VAR$_K(\ell)$ for a lag length of $\ell = 1$ is then given by
\begin{equation} \label{eq:MF-VAR}
    \begin{pmatrix}
        \mathbf{y}(t) \\
        \mathbf{x}^{m_1}(t) \\  
        \vdots \\
        \mathbf{x}^{m_d}(t)
    \end{pmatrix}
    = 
    \mathbf{B}
    \times
    \begin{pmatrix}
        \mathbf{y}(t-1) \\
        \mathbf{x}^{m_1}(t-1) \\ 
        \vdots \\
        \mathbf{x}^{m_d}(t-1)
    \end{pmatrix}
    + 
        \begin{pmatrix}
        \mathbf{u}(t) \\
        \mathbf{u}^{m_1}(t) \\  
        \vdots \\
        \mathbf{u}^{m_d}(t)
    \end{pmatrix},
\end{equation}
where $\mathbf{B} \in \mathbb{R}^{K \times K}$ 
denotes the autoregressive coefficient matrix at lag $1$ for which the total number of 
series $K$ is given by 
$
    K = k_L + m_1\cdot k_1 + \dots +m_d\cdot k_d.
$
Further, $\{\mathbf{u}_t \in \mathbb{R}^{K}\}_{t=1}^T$ is a mean zero error vector with nonsingular contemporaneous covariance matrix $\boldsymbol\Sigma_u$. We focus on stationary time series and assume that all 
series are mean-centered such that no intercept is included.
As for the standard VAR, the $ij^{th}$ entry of $\mathbf{B}$, denoted by $\beta_{ij}$, explains the lagged effect of the $j$th series on the $i$th series.
The entries of the autoregressive coefficient matrix thus permits one to study one-step-ahead predictive Granger causality, an often investigated feature also in the context of MF-VARs (e.g., \citealp{ghysels2016testing}; \citealp{gotz2016testing}). Besides, given the mixed-frequency nature of the variables, it may be of interest to analyze whether knowing the values of the high-frequency variable at time $t$ helps to predict the low-frequency variable during the same time period and vice versa. This form of hidden, contemporaneous Granger causality between high- and low-frequency variables can be derived from the error covariance matrix $\boldsymbol\Sigma_u$ 
(see Section \ref{nowcast restrictions}).

We focus on first-order mixed-frequency VARs  throughout the paper, \revision{extensions to higher-order systems can be easily made, see Remark \ref{remark multiple lags}}. If the true model is multivariate, each component series follows an ARMA$(K\ell, (K-1)\ell)$, see the final equation representation of VARs in \citealp{zellner1974time}, or \citealp{cubadda2009studying} for a more recent discussion. 
Besides, a VAR(1) can even yield long-memory processes for $K \rightarrow \infty$ (\citealp{chevillon2018generating}). 
Hence, as a large system with a small lag length can generate smaller systems with rich and versatile dynamics,
it is plausible to assume that the data generating process of a high-dimensional MF-VAR has a small $\ell$. 
But even a MF-VAR$_K(1)$ model requires one to estimate $K^2$ parameters which quickly becomes large since for a fixed $T$ the parameter vector grows (quadratically) with the number of time series included but also due to the high-per-low frequency observations $m_1, \dots, m_d$. 
In the remainder, we use matrix notation to compactly express the mixed-frequency VAR. To this end, define
\begin{alignat*}{4}
     \bar{\mathbf{y}}_t &= [\mathbf{y}(t)^\prime,
        \mathbf{x}^{m_1}(t)^\prime,
        \dots,
        \mathbf{x}^{m_d}(t)^\prime]^{\prime} \ &&(K \times 1) \qquad &
    \mathbf{Y} &= [\bar{\mathbf{y}}_1,\dots,\bar{\mathbf{y}}_N]^{\prime} \ &&(N \times K) \\
    \mathbf{Z} &= [\bar{\mathbf{y}}_0,\dots,\bar{\mathbf{y}}_{N-1}]^{\prime} \ &&(N \times K) \quad &\mathbf{X} &= \mathbf{I}_K \otimes \mathbf{Z} \ &&(NK \times K^2) \\
     \mathbf{u}_t &= [\mathbf{u}(t)^\prime,  \mathbf{u}^{m_1}(t)^\prime, \dots, \mathbf{u}^{m_d}(t)^\prime]^{\prime} \ &&(K \times 1) \quad
     &\mathbf{U} &= [\mathbf{u}_1,\dots,\mathbf{u}_N]^{\prime} \ &&(N \times K),
\end{alignat*}
where $N = T-1$ are the number of time points  available given the MF-VAR$_K(1)$, 
$\mathbf{I}_K$ denotes the identity matrix of dimension $K$ and $\otimes$  the Kronecker product.
Then the MF-VAR$_K(1)$ can  be written as $\mathbf{y} = \mathbf{X}\bm{\beta} + \mathbf{u}$, with $\mathbf{y}=\text{vec}(\mathbf{Y})$, $\bm{\beta} = \text{vec}(\mathbf{B}^\prime)$, and $\mathbf{u} = \text{vec}(\mathbf{U})$.

In the classical low-dimensional setting $K<N$, the MF-VAR  can be estimated by least squares. However, as the number of parameters grows relative to the time series length $T$,  least squares  becomes unreliable as it results in high variance, overfitting and poor out-of-sample forecasting. We therefore resort to penalization methods. Many authors have used the lasso (\citealp{tibshirani1996regression}), which attains so called ``patternless" sparsity in the parameter matrices of the VAR (e.g., \citealp{hsu2008subset, Demirer17}).
We, instead, use the dynamic structure of the MF-VAR  as a guide in our sparse estimation procedure.
Therefore, we propose regularized estimation of MF-VARs that translates information about the hierarchical structure of low- and high-frequency variables into a convex regularizer that delivers structured sparsity patterns appropriate to the context of MF-VARs. 

\subsection{Hierarchical Structures} \label{section hierarchical structures}
\begin{figure}[t]
    \centering
\begin{tikzpicture}[x=1pt,y=1pt,yscale=-1,xscale=1]

\draw  [color={rgb, 255:red, 16; green, 67; blue, 210  }  ,draw opacity=1 ] (20,20) -- (30,20) -- (30,30) -- (20,30) -- cycle ;
\draw  [color={rgb, 255:red, 16; green, 67; blue, 210  }  ,draw opacity=1 ] (31,31) -- (51,31) -- (51,51) -- (31,51) -- cycle ;
\draw  [color={rgb, 255:red, 208; green, 2; blue, 27 }  ,draw opacity=1 ] (30,31) -- (30,51) -- (20,51) -- (20,31) -- cycle ;
\draw  [color={rgb, 255:red, 126; green, 211; blue, 33 }  ,draw opacity=1 ] (31,20) -- (51,20) -- (51,30) -- (31,30) -- cycle ;
\draw  [color={rgb, 255:red, 16; green, 67; blue, 210  }  ,draw opacity=1 ] (101.5,101.5) -- (141.5,101.5) -- (141.5,141.5) -- (101.5,141.5) -- cycle ;
\draw  [color={rgb, 255:red, 208; green, 2; blue, 27 }  ,draw opacity=1 ] (30,70.5) -- (30,100.5) -- (20,100.5) -- (20,70.5) -- cycle ;
\draw  [color={rgb, 255:red, 208; green, 2; blue, 27 }  ,draw opacity=1 ] (51,70.5) -- (51,100.5) -- (31,100.5) -- (31,70.5) -- cycle ;
\draw  [color={rgb, 255:red, 208; green, 2; blue, 27 }  ,draw opacity=1 ] (100.5,101.5) -- (100.5,141.5) -- (70.5,141.5) -- (70.5,101.5) -- cycle ;
\draw  [color={rgb, 255:red, 126; green, 211; blue, 33 }  ,draw opacity=1 ] (101.5,20) -- (141.5,20) -- (141.5,30) -- (101.5,30) -- cycle ;
\draw  [color={rgb, 255:red, 126; green, 211; blue, 33 }  ,draw opacity=1 ] (101.5,31.11) -- (141.5,31.11) -- (141.5,51.11) -- (101.5,51.11) -- cycle ;
\draw  [color={rgb, 255:red, 126; green, 211; blue, 33 }  ,draw opacity=1 ] (101.5,70.5) -- (141.5,70.5) -- (141.5,100.5) -- (101.5,100.5) -- cycle ;
\draw  [color={rgb, 255:red, 16; green, 67; blue, 210  }  ,draw opacity=1 ] (70.5,70.5) -- (100.5,70.5) -- (100.5,100.5) -- (70.5,100.5) -- cycle ;
\draw  [color={rgb, 255:red, 126; green, 211; blue, 33 }  ,draw opacity=1 ] (70.5,20) -- (100.5,20) -- (100.5,30) -- (70.5,30) -- cycle ;
\draw  [color={rgb, 255:red, 126; green, 211; blue, 33 }  ,draw opacity=1 ] (70.5,31.11) -- (100.5,31.11) -- (100.5,51.11) -- (70.5,51.11) -- cycle ;
\draw  [color={rgb, 255:red, 208; green, 2; blue, 27 }  ,draw opacity=1 ] (30,101.5) -- (30,141.5) -- (20,141.5) -- (20,101.5) -- cycle ;
\draw  [color={rgb, 255:red, 208; green, 2; blue, 27 }  ,draw opacity=1 ] (51,101.5) -- (51,141.5) -- (31,141.5) -- (31,101.5) -- cycle ;
\draw   (130,14) -- (147,14) -- (147,146.83) ;
\draw   (130,146.83) -- (147,146.83) -- (147,125.71) ;

\draw   (31,146.83) -- (14,146.83) -- (14,14) ;
\draw   (31,14) -- (14,14) -- (14,35.12) ;

\draw    (159.86,0.74) -- (159.57,160.17) ;
\draw  [draw opacity=0] (259.86,60) -- (360.36,60) -- (360.36,120.5) -- (259.86,120.5) -- cycle ; \draw  [color={rgb, 255:red, 126; green, 211; blue, 33 }  ,draw opacity=1 ] (259.86,60) -- (259.86,120.5)(279.86,60) -- (279.86,120.5)(299.86,60) -- (299.86,120.5)(319.86,60) -- (319.86,120.5)(339.86,60) -- (339.86,120.5)(359.86,60) -- (359.86,120.5) ; \draw  [color={rgb, 255:red, 126; green, 211; blue, 33 }  ,draw opacity=1 ] (259.86,60) -- (360.36,60)(259.86,80) -- (360.36,80)(259.86,100) -- (360.36,100)(259.86,120) -- (360.36,120) ; \draw  [color={rgb, 255:red, 126; green, 211; blue, 33 }  ,draw opacity=1 ]  ;
\draw  [draw opacity=0] (369.86,20) -- (430.36,20) -- (430.36,120.5) -- (369.86,120.5) -- cycle ; \draw  [color={rgb, 255:red, 208; green, 2; blue, 27 }  ,draw opacity=1 ] (369.86,20) -- (369.86,120.5)(389.86,20) -- (389.86,120.5)(409.86,20) -- (409.86,120.5)(429.86,20) -- (429.86,120.5) ; \draw  [color={rgb, 255:red, 208; green, 2; blue, 27 }  ,draw opacity=1 ] (369.86,20) -- (430.36,20)(369.86,40) -- (430.36,40)(369.86,60) -- (430.36,60)(369.86,80) -- (430.36,80)(369.86,100) -- (430.36,100)(369.86,120) -- (430.36,120) ; \draw  [color={rgb, 255:red, 208; green, 2; blue, 27 }  ,draw opacity=1 ]  ;
\draw  [draw opacity=0] (169.86,40) -- (250.36,40) -- (250.36,120.5) -- (169.86,120.5) -- cycle ; \draw  [color={rgb, 255:red, 16; green, 67; blue, 210  }  ,draw opacity=1 ] (169.86,40) -- (169.86,120.5)(189.86,40) -- (189.86,120.5)(209.86,40) -- (209.86,120.5)(229.86,40) -- (229.86,120.5)(249.86,40) -- (249.86,120.5) ; \draw  [color={rgb, 255:red, 16; green, 67; blue, 210  }  ,draw opacity=1 ] (169.86,40) -- (250.36,40)(169.86,60) -- (250.36,60)(169.86,80) -- (250.36,80)(169.86,100) -- (250.36,100)(169.86,120) -- (250.36,120) ; \draw  [color={rgb, 255:red, 16; green, 67; blue, 210  }  ,draw opacity=1 ]  ;

\draw (57,55) node [anchor=north west][inner sep=0.75pt]  [color={rgb, 255:red, 16; green, 67; blue, 210  }  ,opacity=1 ,rotate=-45] [align=left] {\dots};
\draw (27,55) node [anchor=north west][inner sep=0.75pt]  [color={rgb, 255:red, 208; green, 2; blue, 27 }  ,opacity=1 ,rotate=-90] [align=left] {\dots};
\draw (37,55) node [anchor=north west][inner sep=0.75pt]  [color={rgb, 255:red, 208; green, 2; blue, 27 }  ,opacity=1 ,rotate=-45] [align=left] {\dots};
\draw (77,55) node [anchor=north west][inner sep=0.75pt]  [color={rgb, 255:red, 126; green, 211; blue, 33 }  ,opacity=1 ,rotate=-45] [align=left] {\dots};
\draw (125,55) node [anchor=north west][inner sep=0.75pt]  [color={rgb, 255:red, 126; green, 211; blue, 33 }  ,opacity=1 ,rotate=-90] [align=left] {\dots};
\draw (57,75) node [anchor=north west][inner sep=0.75pt]  [color={rgb, 255:red, 208; green, 2; blue, 27 }  ,opacity=1 ,rotate=-45] [align=left] {\dots};
\draw (57,35) node [anchor=north west][inner sep=0.75pt]  [color={rgb, 255:red, 126; green, 211; blue, 33 }  ,opacity=1 ,rotate=-45] [align=left] {\dots};
\draw (67,128) node [anchor=north west][inner sep=0.75pt]  [color={rgb, 255:red, 208; green, 2; blue, 27 }  ,opacity=1 ,rotate=-180] [align=left] {\dots};
\draw (67,26) node [anchor=north west][inner sep=0.75pt]  [color={rgb, 255:red, 126; green, 211; blue, 33 }  ,opacity=1 ,rotate=-180] [align=left] {\dots};

\draw (210.6,132.6) node  [font=\footnotesize] [align=left] {\begin{minipage}[lt]{50.7875pt}\setlength\topsep{0pt}
\begin{center}
Own-on-Own
\end{center}

\end{minipage}};
\draw (311,132.6) node  [font=\footnotesize] [align=left] {\begin{minipage}[lt]{48.991875pt}\setlength\topsep{0pt}
\begin{center}
Higher-on-Lower
\end{center}

\end{minipage}};
\draw (401.8,132.6) node  [font=\footnotesize] [align=left] {\begin{minipage}[lt]{48.991875pt}\setlength\topsep{0pt}
\begin{center}
Lower-on-Higher
\end{center}

\end{minipage}};
\draw (210.31,148.8) node    {$m \times m$};
\draw (311.91,148.8) node    {$m_{L} \times m_{H}$};
\draw (403.11,148.8) node    {$m_{H} \times m_{L}$};
\end{tikzpicture}
    \caption{Division of autoregressive coefficient matrix $\mathbf{B}$ into sub-matrices capturing Own-on-Own, Higher-on-Lower and Lower-on-Higher effects.}
    \label{fig:division coefficient matrix}
\end{figure}
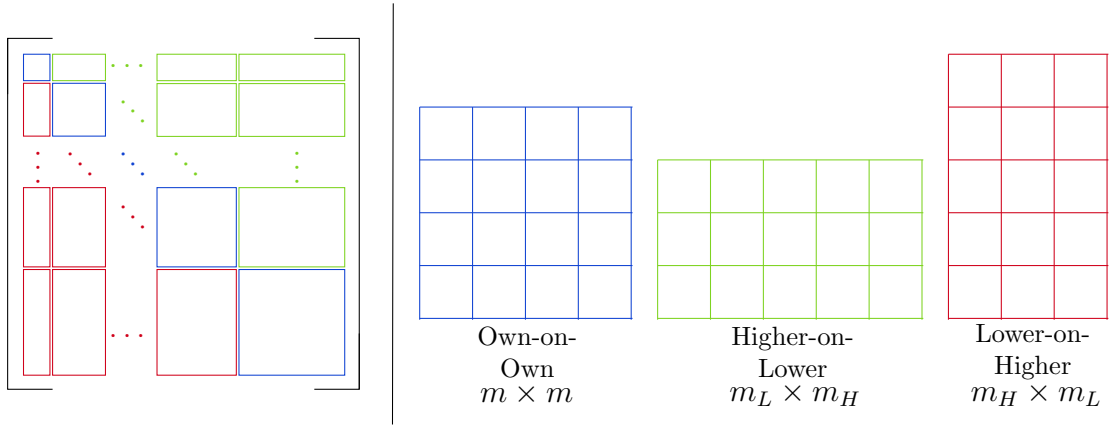

We describe the hierarchical sparsity patterns that arise in the autoregressive coefficient matrix $\mathbf{B}$ of a MF-VAR. We demonstrate the intuition,  for ease of notation,  with $k_L=k_1=\dots=k_d=1$,
see Remark \ref{remark multiple series} for an extension to multiple series.
The parameters of the matrix $\mathbf{B}$ can be divided in $(d+1)^2$ different groups as depicted by the sub-matrices in the left panel of Figure \ref{fig:division coefficient matrix}. We distinguish three types of sub-matrices capturing respectively \textit{Own-on-Own}, \textit{Higher-on-Lower} and \textit{Lower-on-Higher}  effects, \revision{to extend standard VAR estimation with single-frequencies to mixed-frequencies}. The \textit{Own-on-Own} sub-matrices are $m\times m$ square matrices that lie on the main diagonal and describe the effects of a series' own lags on itself (with $m_L=1$ for the low-frequency variable).
The \textit{Higher-on-Lower} sub-matrices are short, wide $m_L \times m_H$ matrices, where $L$ and $H$ refer to the corresponding lower- and higher-frequency variable. They lie above the main diagonal and contain the lagged effect of a higher-frequency series onto a series with respective lower frequency. 
Note that $L$ (and $H$) just identify which of the two variables is of lower (and higher) frequency.
Thus, this group does not only contain the effects of the higher-frequency variables onto the variable with the lowest frequency but also describes the interactions between the higher-frequency variables. For instance, in a quarter/month/week-example, these incorporate the effects of the monthly and weekly variable onto the quarterly variable but also the effects of the weekly onto the monthly variable.
The \textit{Lower-on-Higher} sub-matrices are long, thin $m_H \times m_L$ matrices. They lie below the main diagonal and contain the lagged effect depicting the effect of a lower-frequency series onto a higher-frequency series.

For each sub-matrix, we impose a hierarchical priority structure for parameter inclusion. Parameters with higher priority within one group should be included in the model before parameters with lower priority within the same group, where the priority value of each parameter depends on how informative a practitioner finds the associated regressor.
A priority value of one indicates highest priority for  inclusion. More precisely, we introduce a priority value $p_g^{ij} \in \{1, \dots, P_g\}$ for each element $ij$ belonging to parameter group $g=1,\dots,G$ in the matrix $\mathbf{B}$ to denote its inclusion priority. 
Here, we order the groups in Figure \ref{fig:division coefficient matrix} from left to right and top to bottom. For instance, the top left square is group $g=1$, the rectangle to the right is group $g=2$.
If $p_g^{ij} < p_g^{i'j'}$ for $i\neq i'$ and $j\neq j'$, we prioritize parameter $\beta_{ij}$ over $\beta_{i'j'}$ in the model. The hierarchical structure is imposed for each group $g$ individually and hence in each group the priority values start with value 1 (highest priority) and go up to $P_g$ (lowest priority). We do not encode structures where certain parameters of one group should enter the model before certain parameters of another group as our aim is to encode priority of parameter inclusion with respect to the effects of one frequency component (Own/Higher/Lower) on another (Own/Higher/Lower). 

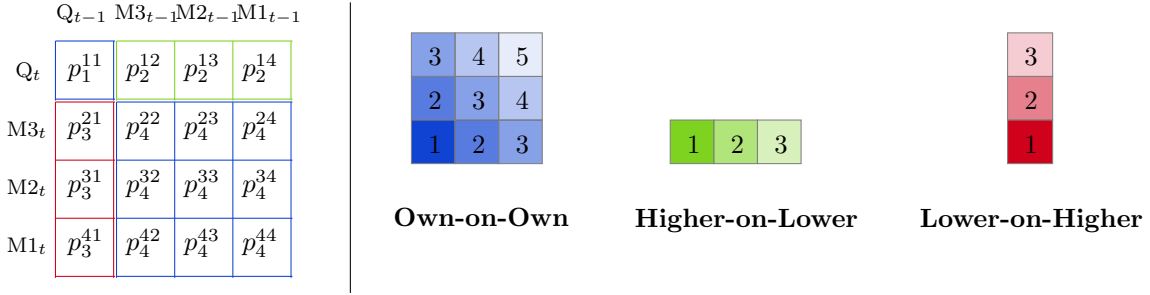
\begin{figure}
    \centering
    \begin{tikzpicture}[x=1.1pt,y=1.1pt,yscale=-1,xscale=1]

\draw  [draw opacity=0] (59.6,44.3) -- (120.1,74.3) -- (120.1,134.8) -- (59.6,134.8) -- cycle ; \draw  [color={rgb, 255:red, 16; green, 67; blue, 210 }  ,draw opacity=1 ] (59.6,34.3) -- (59.6,94.8)(79.6,34.3) -- (79.6,94.8)(99.6,34.3) -- (99.6,94.8)(119.6,34.3) -- (119.6,94.8) ; \draw  [color={rgb, 255:red, 16; green, 67; blue, 210 }  ,draw opacity=1 ] (59.6,34.3) -- (120.1,34.3)(59.6,54.3) -- (120.1,54.3)(59.6,74.3) -- (120.1,74.3)(59.6,94.3) -- (120.1,94.3) ; \draw  [color={rgb, 255:red, 16; green, 67; blue, 210 }  ,draw opacity=1 ]  ;
\draw  [draw opacity=0] (38.6,34.3) -- (59.1,34.3) -- (59.1,94.8) -- (38.6,94.8) -- cycle ; \draw  [color={rgb, 255:red, 208; green, 2; blue, 27 }  ,draw opacity=1 ] (38.6,34.3) -- (38.6,94.8)(58.6,34.3) -- (58.6,94.8) ; \draw  [color={rgb, 255:red, 208; green, 2; blue, 27 }  ,draw opacity=1 ] (38.6,34.3) -- (59.1,34.3)(38.6,54.3) -- (59.1,54.3)(38.6,74.3) -- (59.1,74.3)(38.6,94.3) -- (59.1,94.3) ; \draw  [color={rgb, 255:red, 208; green, 2; blue, 27 }  ,draw opacity=1 ]  ;
\draw  [draw opacity=0] (59.6,53.3) -- (120.1,53.3) -- (120.1,73.8) -- (59.6,73.8) -- cycle ; \draw  [color={rgb, 255:red, 126; green, 211; blue, 33 }  ,draw opacity=1 ] (59.6,13.3) -- (59.6,33.8)(79.6,13.3) -- (79.6,33.8)(99.6,13.3) -- (99.6,33.8)(119.6,13.3) -- (119.6,33.8) ; \draw  [color={rgb, 255:red, 126; green, 211; blue, 33 }  ,draw opacity=1 ] (59.6,13.3) -- (120.1,13.3)(59.6,33.3) -- (120.1,33.3) ; \draw  [color={rgb, 255:red, 126; green, 211; blue, 33 }  ,draw opacity=1 ]  ;
\draw  [draw opacity=0] (38.6,53.3) -- (59.1,53.3) -- (59.1,73.8) -- (38.6,73.8) -- cycle ; \draw  [color={rgb, 255:red, 16; green, 67; blue, 210 }  ,draw opacity=1 ] (38.6,13.3) -- (38.6,33.8)(58.6,13.3) -- (58.6,33.8) ; \draw  [color={rgb, 255:red, 16; green, 67; blue, 210 }  ,draw opacity=1 ] (38.6,13.3) -- (59.1,13.3)(38.6,33.3) -- (59.1,33.3) ; \draw  [color={rgb, 255:red, 16; green, 67; blue, 210 }  ,draw opacity=1 ]  ;

\filldraw[fill={rgb, 255:red, 208; green, 2; blue, 27 }, fill opacity=0.2, draw=gray] (366,10.5) rectangle (381.5,25.5);
\filldraw[fill={rgb, 255:red, 208; green, 2; blue, 27 }, fill opacity=0.5, draw=gray] (366,25.5) rectangle (381.5,40.5);
\filldraw[fill={rgb, 255:red, 208; green, 2; blue, 27 }, fill opacity=1, draw=gray] (366,40.5) rectangle (381.5,55.5);

\filldraw[fill={rgb, 255:red, 126; green, 211; blue, 33}, fill opacity=1, draw=gray] (250,40.5) rectangle (265,55.5);
\filldraw[fill={rgb, 255:red, 126; green, 211; blue, 33}, fill opacity=0.6, draw=gray] (265,40.5) rectangle (280,55.5);
\filldraw[fill={rgb, 255:red, 126; green, 211; blue, 33}, fill opacity=0.3, draw=gray] (280,40.5) rectangle (295,55.5);

\draw    (140,0) -- (140,100) ;

\filldraw[fill={rgb, 255:red, 16; green, 67; blue, 210 }  ,fill opacity=0.5, draw=gray] (161,10.5) rectangle (176,25.5);
\filldraw[fill={rgb, 255:red, 16; green, 67; blue, 210 }  ,fill opacity=0.3, draw=gray] (176,10.5) rectangle (191,25.5);
\filldraw[fill={rgb, 255:red, 16; green, 67; blue, 210 }  ,fill opacity=0.1, draw=gray] (191,10.5) rectangle (206,25.5);
\filldraw[fill={rgb, 255:red, 16; green, 67; blue, 210 }  ,fill opacity=0.7, draw=gray] (161,25.5) rectangle (176,40.5);
\filldraw[fill={rgb, 255:red, 16; green, 67; blue, 210 }  ,fill opacity=0.5, draw=gray] (176,25.5) rectangle (191,40.5);
\filldraw[fill={rgb, 255:red, 16; green, 67; blue, 210 }  ,fill opacity=0.3, draw=gray] (191,25.5) rectangle (206,40.5);
\filldraw[fill={rgb, 255:red, 16; green, 67; blue, 210 }  ,fill opacity=1, draw=gray] (161,40.5) rectangle (176,55.5);
\filldraw[fill={rgb, 255:red, 16; green, 67; blue, 210 }  ,fill opacity=0.7, draw=gray] (176,40.5) rectangle (191,55.5);
\filldraw[fill={rgb, 255:red, 16; green, 67; blue, 210 }  ,fill opacity=0.5, draw=gray] (191,40.5) rectangle (206,55.5);

\draw (154,70) node [anchor=north west][inner sep=0.75pt]   [align=left] {{\footnotesize \textbf{Own-on-Own}}};
\draw (237,70) node [anchor=north west][inner sep=0.75pt]   [align=left] {{\footnotesize \textbf{Higher-on-Lower}}};
\draw (335,70) node [anchor=north west][inner sep=0.75pt]   [align=left] {{\footnotesize \textbf{Lower-on-Higher}}};
\draw (166,45) node [anchor=north west][inner sep=0.75pt]  [font=\footnotesize]  {$1$};
\draw (166,30) node [anchor=north west][inner sep=0.75pt]  [font=\footnotesize]  {$2$};
\draw (181,45) node [anchor=north west][inner sep=0.75pt]  [font=\footnotesize]  {$2$};
\draw (166,15) node [anchor=north west][inner sep=0.75pt]  [font=\footnotesize]  {$3$};
\draw (181,30) node [anchor=north west][inner sep=0.75pt]  [font=\footnotesize]  {$3$};
\draw (196,45) node [anchor=north west][inner sep=0.75pt]  [font=\footnotesize]  {$3$};
\draw (181,15) node [anchor=north west][inner sep=0.75pt]  [font=\footnotesize]  {$4$};
\draw (196,30) node [anchor=north west][inner sep=0.75pt]  [font=\footnotesize]  {$4$};
\draw (196,15) node [anchor=north west][inner sep=0.75pt]  [font=\footnotesize]  {$5$};

\draw (255,45) node [anchor=north west][inner sep=0.75pt]  [font=\footnotesize]  {$1$};
\draw (270,45) node [anchor=north west][inner sep=0.75pt]  [font=\footnotesize]  {$2$};
\draw (285,45) node [anchor=north west][inner sep=0.75pt]  [font=\footnotesize]  {$3$};
\draw (24,20) node [anchor=north west][inner sep=0.75pt]  [font=\scriptsize]  {Q$_{t}$};
\draw (21,40) node [anchor=north west][inner sep=0.75pt]  [font=\scriptsize]  {M3$_{t}$};
\draw (21,60) node [anchor=north west][inner sep=0.75pt]  [font=\scriptsize]  {M2$_{t}$};
\draw (21,80) node [anchor=north west][inner sep=0.75pt]  [font=\scriptsize]  {M1$_{t}$};
\draw (42,17) node [anchor=north west][inner sep=0.75pt]  [font=\footnotesize]  {$p^{11}_{1}$};
\draw (62,17) node [anchor=north west][inner sep=0.75pt]  [font=\footnotesize]  {$p^{12}_{2}$};
\draw (82,17) node [anchor=north west][inner sep=0.75pt]  [font=\footnotesize]  {$p^{13}_{2}$};
\draw (102,17) node [anchor=north west][inner sep=0.75pt]  [font=\footnotesize]  {$p^{14}_{2}$};
\draw (42,37) node [anchor=north west][inner sep=0.75pt]  [font=\footnotesize]  {$p^{21}_{3}$};
\draw (42,57) node [anchor=north west][inner sep=0.75pt]  [font=\footnotesize]  {$p^{31}_{3}$};
\draw (42,77) node [anchor=north west][inner sep=0.75pt]  [font=\footnotesize]  {$p^{41}_{3}$};
\draw (62,37) node [anchor=north west][inner sep=0.75pt]  [font=\footnotesize]  {$p^{22}_{4}$};
\draw (82,37) node [anchor=north west][inner sep=0.75pt]  [font=\footnotesize]  {$p^{23}_{4}$};
\draw (102,37) node [anchor=north west][inner sep=0.75pt]  [font=\footnotesize]  {$p^{24}_{4}$};
\draw (62,57) node [anchor=north west][inner sep=0.75pt]  [font=\footnotesize]  {$p^{32}_{4}$};
\draw (82,57) node [anchor=north west][inner sep=0.75pt]  [font=\footnotesize]  {$p^{33}_{4}$};
\draw (102,57) node [anchor=north west][inner sep=0.75pt]  [font=\footnotesize]  {$p^{34}_{4}$};
\draw (62,77) node [anchor=north west][inner sep=0.75pt]  [font=\footnotesize]  {$p^{42}_{4}$};
\draw (82,77) node [anchor=north west][inner sep=0.75pt]  [font=\footnotesize]  {$p^{43}_{4}$};
\draw (102,77) node [anchor=north west][inner sep=0.75pt]  [font=\footnotesize]  {$p^{44}_{4}$};
\draw (38,0) node [anchor=north west][inner sep=0.75pt]  [font=\scriptsize]  {Q$_{t-1}$};
\draw (58,0) node [anchor=north west][inner sep=0.75pt]  [font=\scriptsize]  {M3$_{t-1}$};
\draw (79,0) node [anchor=north west][inner sep=0.75pt]  [font=\scriptsize]  {M2$_{t-1}$};
\draw (100,0) node [anchor=north west][inner sep=0.75pt] {\scriptsize M1$_{t-1}$};
\draw (371,45) node [anchor=north west][inner sep=0.75pt]  [font=\footnotesize]  {$1$};
\draw (371,30) node [anchor=north west][inner sep=0.75pt]  [font=\footnotesize]  {$2$};
\draw (371,15) node [anchor=north west][inner sep=0.75pt]  [font=\footnotesize]  {$3$};
\end{tikzpicture}
\vspace{-1cm}
    \caption{Left: Priority values $p^{ij}_g$, corresponding to each element $ij$ in the autoregressive coefficient matrix $\mathbf{B}$ for a MF-VAR$_4(1)$. Right: Recency-based penalty structures for parameter groups Own-on-Own, Higher-on-Lower and Lower-on-Higher in a MF-VAR$_4(1)$. }
    \label{fig:HIER_struc 4(1)}
\end{figure}

These hierarchical priority structures are highly general and could potentially accommodate various
structured sparsity patterns in the AR parameter matrix that researchers or practitioners  want to encourage.
\revision{We give special attention to a recency-based priority structure,} 
where the priority value $p_g^{ij}$ is set according to the recency of the information the $j$th time series of $\bar{\mathbf{y}}_{t-1}$ contains relative to the $i$th series of $\bar{\mathbf{y}}_t$. The more recent the information contained in the lagged predictor, the more informative, and thus the  higher its inclusion priority. 
Figure \ref{fig:HIER_struc 4(1)} visualizes the recency-based structure 
for 
an example with only one quarterly and one monthly variable having a $4\times 4$ coefficient matrix $\textbf{B}$ and $G=4$ groups. 
For instance, consider the Higher-on-Lower parameter block, where month three of the previous quarter contains the most recent information, followed by month two and then month one. 
The priority values are thus increasing from left to right.

\revision{The recency-based priority structure is conceptually similar to other often used restriction schemes in the (MF)-VAR literature.}
For the Higher-on-Lower group, it is similar to  a MIDAS weighting function with decaying shape, for instance an exponential Almon lag polynomial. Both approaches assume a decaying memory pattern in the economic processes, however, in our setting, we do not restrict the parameters to a specific nonlinear function. 
\revision{Besides, our recency-based priorities are similar in spirit to the Minnesota prior in the Bayesian literature (e.g., \citealp{cimadomo2021nowcasting}) in the sense that longer lags are treated differently than shorter lags as}
\revision{they are expected to contain less information about a variable's current value.} 

\begin{remark} \label{remark multiple series}
If one were to extend to a MF-VAR$_K(1)$ with multiple time series per frequency component, the total number of groups $G$ becomes $(k_L+k_1+\dots+k_d)^2$. Then the priority values describing the dependence of the dependent variable having frequency $m_i$ on the independent variable having frequency $m_j$ would be replicated $k_{i} \times k_{j}$ times.
E.g., in a setting with two quarterly and three monthly variables, the priority values describing the effect of monthly series on quarterly series are replicated six times.
\end{remark}

\begin{remark} \label{remark multiple lags}
\revision{
If one were to increase the lag length, the total number of groups $G$ would remain unchanged, but
each group would additionally
incorporate all its
higher-order lagged coefficients. 
Since higher-order lags contain older information, we decrease their priority of inclusion. 
For example,
the Higher-on-Lower group for a MF-VAR$_4(2)$ would have priority values one up to three for lag 1; four up to six for lag 2. We discuss the sensitivity of our empirical results to the maximum lag length in Section \ref{Macroeconomic Application Sensitivity Analysis}.}
\end{remark}

\subsection{Nowcasting Restrictions} \label{nowcast restrictions}
The hidden contemporaneous links between high- and low-frequency variables can be investigated in the covariance matrix ${\boldsymbol\Sigma}_{u}$ of the error terms ${\bf u}_t$ in the MF-VAR model \eqref{eq:MF-VAR}.  \cite{gotz2014nowcasting} test for block diagonality of ${\boldsymbol\Sigma}_{u}$ to investigate the null of no contemporaneous Granger causality or, to put it differently, the absence of nowcasting relationships between high- and low-frequency indicators. In the remainder, we refer to contemporaneous Granger causality  as ``nowcasting causality". The authors show that the conditional single equation model (e.g., MIDAS) with contemporaneous regressors can be misleading as it can change the dynamics observed in a MF-VAR system (e.g., Granger causality relations). We do not face this problem since we work with the reduced form MF-VAR and not with a single equation conditional model derived from the MF-VAR. 

In the context of our paper, a detailed inspection of the residual covariance matrix $\widehat{\boldsymbol\Sigma}_{u}:= \frac{1}{T-\ell}\sum_{t = \ell+1}^{T} \widehat{{\bf u}}_t \widehat{{\bf u}}^\prime_t$ of the high-dimensional MF-VAR is interesting  for (at least) two reasons. 
First, from an economic perspective, we aim to investigate whether there exist high-frequency months, weeks or days of some series that nowcast low-frequency variables. 
Since this is a correlation measure observed in the symmetric blocks of the residual covariance matrix, we obviously cannot point towards a direction of this contemporaneous link. \citeauthor{lutkepohl2005new} (\citeyear{lutkepohl2005new}, pages 45-48) stresses that the \textquotedblleft \textit{direction of (nowcasting) causation must be obtained from further knowledge (e.g., economic theory) on the relationship between the variables}\textquotedblright. We can only agree on that. Second, from a statistical perspective,  we aim to compare the performance of the MF-VAR from Section \ref{section hierarchical structures} when additional restrictions on the error covariance matrix are considered. Intuitively, this is a Generalized Least Squares (GLS) type improvement over the ``unrestricted" hierarchical MF-VAR. 

In our economic application (to be discussed in Section \ref{Macroeconomic Application}), we consider quarterly real GDP growth as one of the main low-frequency indicators, and are interested in investigating whether some high-frequency monthly (e.g., retail sales) and/or weekly (e.g., money stock, federal fund rate) series can deliver a coincident indicator of GDP growth with the advantage being that they are released earlier. We do not claim that they directly impact GDP but simply that the business cycle movements detected in those high-frequency variables track the fluctuations of the GDP growth well.

To investigate such nowcasting relations, let us assume a single low-frequency variable and decompose ${\boldsymbol\Sigma}_{u}$ into four blocks
\[
{\boldsymbol\Sigma}_{u}=\left[ 
\begin{array}{cc}
\sigma _{1.1}^{2} & {\boldsymbol \sigma}_{.K-1}^{2^{\prime }} \\ 
{\boldsymbol \sigma}_{.K-1}^{2} & {\boldsymbol\Sigma}_{2:K}
\end{array}
\right], 
\]
with ${\boldsymbol\Sigma}_{2:K}$ the $(K-1)\times (K-1)$ block of ${\boldsymbol\Sigma}_{u}$ corresponding to the covariances between errors of the high-frequency variables. It contains both (co)-variances per high-frequency variable (i.e.,\ its main-diagonal blocks) as well as covariances between different high-frequency variables (i.e.,\ its off-diagonal blocks). $\sigma^2_{1.1}$ is the error variance of the low-frequency variables. It is a scalar when there is only one variable but in general it is a square matrix. ${\boldsymbol \sigma}_{.K-1}^2$ is a vector when there is a single low-frequency variable, with the cross-covariances of the errors between low- and high-frequencies. This will be the block we focus on  to detect nowcasting relations between each low-frequency variable and each high-frequency variable.

To detect nowcasting relations, we investigate the existence of a sparse matrix \[
{\boldsymbol\Sigma}_{u}^{\ast}=\left[ 
\begin{array}{cc}
\sigma_{1.1}^{\ast 2} & {\boldsymbol \sigma}_{.K-1}^{\ast 2^{\prime }} \\ 
{\boldsymbol \sigma}_{.K-1}^{\ast 2} & {\boldsymbol\Sigma}_{2:K}^{\ast }
\end{array}
\right],
\]
where we leave $\sigma_{1.1}^{\ast 2}$ as well as the 
main-diagonal blocks in ${\boldsymbol\Sigma}_{2:K}^{\ast }$ unrestricted.
The sparsity of the block ${\boldsymbol \sigma}_{.K-1}^{\ast 2}$ is our main focus as it allows us to detect nowcasting relations between the low-frequency variables and each of the high-frequency ones.
The sparsity in the off-diagonal blocks of ${\boldsymbol\Sigma}_{2:K}^{\ast }$ we allow for is not of main economic interest to our nowcasting application, 
but it may be of interest in other applications and it does facilitate retrieval of a positive definite ${\boldsymbol\Sigma}_{u}^{\ast }$ matrix (see Section 3.2). 

\begin{remark}
The case ${\boldsymbol \sigma}_{.K-1}^{\ast 2}={\bf 0}_{(K-1)\times 1}$ (with one low-frequency
variable) corresponds to the block diagonality of ${\boldsymbol\Sigma}_{u}^{\ast }$ and
hence the null of no nowcasting causality.
\end{remark}

\begin{remark}
Unlike for the autoregressive parameter matrix, we do not impose a hierarchical sparsity structure on ${\boldsymbol \sigma}_{.K-1}^{\ast 2}$
because the middle month could be a better coincident indicator for a
quarterly variable than the last month, for instance. Since this is an empirical issue, we prefer to stick to a regularized estimator of the residual covariance matrix that encourages ``patternless" sparsity in $\widehat{{\boldsymbol \sigma}}_{.K-1}^{\ast 2}$.
\end{remark}

Finally, to build our coincident indicator, 
we use a simple procedure which consists of first selecting the variable-period combinations corresponding to non-zero entries in ${\boldsymbol \sigma}_{.K-1}^{\ast 2}$, standardizing the series and subsequently constructing the first principal component. We rescale the coincident indicator to GDP growth 
as in \cite{lewis2020us}.

\section{Regularized Estimation Procedure of MF-VARs} \label{Regularized Estimation Procedure of MF-VARs}

\subsection{Hierarchical Group Lasso for Structured MF-VAR} \label{subsection Hierarchical Group Lasso for Structured MF-VAR} 
To attain the hierarchical structure presented in Section \ref{section hierarchical structures}, we use a nested group lasso (\citealp{zhao2009composite}) which has been successfully employed for various statistical problems, among which, time series models (e.g., \citealp{nicholson2020high}), generalized additive models (e.g., \citealp{lou2016sparse}), regression models with interactions (e.g., \citealp{haris2016convex}) or banded covariance estimation (e.g., \citealp{bien2016convex}).
The group lasso uses the sum of (unsquared) Euclidean norms as penalty term to encourage sparsity on the group-level. Then, either all parameters of a group are set to zero or none. 
By using nested groups, hierarchical sparsity constraints are imposed where one set of parameters being zero implies that another set is also set to zero. 
We encourage hierarchical sparsity within each group of parameters by prioritizing parameters with a lower priority value over parameters with a higher 
one. The proposed hierarchical group estimator for the MF-VAR 
is given by
\begin{align} \label{eq:optimization problem}
    \widehat{\bm{\beta}}&= \underset{\bm{\beta}}{\text{argmin}} \ 
    \Big\{\frac{1}{2}\lVert \mathbf{y} - \mathbf{X} \bm{\beta} \rVert_2^2 + \lambda_{\beta} \mathcal{P}_{\text{Hier}}(\bm{\beta}) \Big \},
\end{align}
where 
$\mathcal{P}_{\text{Hier}}(\bm{\beta})$ denotes the hierarchical group penalty and $\lambda_\beta \geq 0$ is a tuning parameter. 

Before introducing the hierarchical group penalty, recall that we distinguish $g =1,\dots, G$ parameter groups in the autoregressive parameter matrix and that $P_g$ is the maximum priority value of parameter group $g$. 
To impose the hierarchical structure within each parameter group $g$, we consider $P_g$ nested sub-groups $s_g^{(1)},\ldots, s^{(P_g)}_g$. Group $s_g^{(1)}$ contains all parameters of group $g$, 
$s_g^{(2)}$ omits those parameters having priority value one, and finally the last sub-group $s_g^{(P_g)}$ only contains those parameters having the highest priority value $P_g$. Clearly, a nested structure arises with $s_g^{(1)} \supset \dots \supset s_g^{(P_g)}$. 
Now, denote $\bm{\beta}_g^{(p:P_g)} = 
[{\bm{\beta}_{g}^{p}}^\prime, \dots, {\bm{\beta}_{g}^{P_g}}^\prime]^\prime$,
for $1 \leq p \leq P_g$, where $\bm{\beta}_{g}^{p}$ collects the parameters of group $g$ having priority value $p$. We are now ready to define the hierarchical penalty function as 
\begin{equation}\label{eq:penaltyhier}
    \mathcal{P}_{\text{Hier}}(\bm{\beta}) =\sum_{g=1}^G \sum_{p = 1}^{P_g} w_{s_g^{(p)}}\lVert  \bm{\beta}_{g}^{(p:P_g)}\rVert_2.
\end{equation} 
The hierarchical structure within each group $g$
is built in through the condition that if $\bm{\beta}_{g}^{(p:P_g)} = \bf{0}$, then $\bm{\beta}_{g}^{(p':P_g)} = \bf{0}$ where $p < p'$. 

\begin{remark}\label{remark:babii}
\revision{A related sparse-group lasso penalty 
has been proposed by \cite{babii2021machine}. Similar to our penalty,
theirs is tailored to the dynamic nature of the mixed-frequency model to account for different high-frequency lags  being temporally related. As such, both penalties induce sparsity at two levels: whether a variable is important for explaining another or not;  
and if important, both  steer the effect's duration via an additional sparsity layer.
Via our nested group lasso, we do this by building a hierarchy to cut off the effect of one variable on another after a certain (high-frequency) lag. Via the sparse-group lasso, \cite{babii2021machine} regulate the shape of the MIDAS weight function for each group which consists of all high-frequency lags of a single variable.}
\end{remark}

The weights $w_{s_g^{(p)}}$ balance unequally sized nested sub-groups. We take 
$w_{s_g^{(p)}}  = \text{card}(s_g^{(1)})-\text{card}(s_g^{(p)})+1,$
where $\text{card}(\cdot)$ corresponds to the cardinality. As the cardinality of the sub-groups $s_g^{(p)}$ is decreasing with $p$, the weights of the nested sub-groups are increasing with $p$. Sub-groups containing parameters with lower priority, i.e., with older information about the state of the economy, are thus penalized more and hence more likely to be zeroed out. 
\begin{remark}\label{Remark:weights}
\revision{A simple alternative to our proposed weights are equal weights as used in \cite{zhao2009composite} or \cite{nicholson2020high}, which would lead to less aggressive shrinkage.
Simulations in
Section \ref{subsection Autoregressive Estimation Accuracy and Variable Selection}
reveal that the proposed more aggressive weights are preferable for sparsity recognition. In Section \ref{subsection Varying Degree of Density}, we include recommendations to guide forecasters in choosing the weights.}
\end{remark}

Finally, we propose a proximal gradient algorithm (see e.g., \citealp{tseng2008accelerated}) to efficiently solve the optimization problem in 
Equation (\ref{eq:optimization problem}), as detailed in Appendix A.

\subsection{Regularized Estimation of Nowcasting Causality Relations}
To detect nowcasting relations between the low- and high-frequency variables, we use a lasso-penalty 
to impose ``patternless" sparsity on the covariances between low- and high-frequency errors and the covariances between different high-frequency errors (see Section \ref{nowcast restrictions}).
The proposed sparse covariance estimator is then given by
\begin{equation}\label{eq:Sigma obj func}
    \widehat{\bm{\Sigma}}^\ast_u = \underset{\bm{\Sigma}\succ 0}{\text{argmin}} \left \{ \frac{1}{2} \left \lVert \widehat{\boldsymbol\Sigma}_{u} - \bm{\Sigma} \right \rVert_F^2 + \lambda_\Sigma \left \lVert \bm{\Sigma^-} \right \rVert_1   \right \},
\end{equation}
where $\lambda_\Sigma \geq 0$ is a tuning parameter, $\bm{\Sigma}^-$ are the elements of the off-diagonal blocks of $\bm{\Sigma}$. Furthermore, for a matrix $\mathbf{A}$, $\lVert \mathbf{A} \rVert_F = \lVert \text{vec}(\mathbf{A})\rVert_F = (\sum_{ij} \mathbf{A}_{ij}^2)^{1/2}$ denotes the Frobenius norm and $\lVert \mathbf{A} \rVert_1 = \lVert \text{vec}(\mathbf{A})\rVert_1 = \sum_{ij} | \mathbf{A}_{ij} |$ the $l_1$-norm. 

\begin{remark}
The mere addition of the $l_1$-penalty in 
problem \eqref{eq:Sigma obj func} does not guarantee the estimator $  \widehat{\bm{\Sigma}}^\ast_u $ to be positive definite (see \citealp{rothman2009generalized}).
To ensure its positive definiteness,  
the constraint $\bm{\Sigma}\succ 0$ implies that we only consider solutions with strictly positive eigenvalues. \cite{rothman2009generalized} show that \eqref{eq:Sigma obj func} without the constraint $\bm{\Sigma}\succ 0$ essentially boils down to element-wise soft-thresholding of $\bm{\Sigma}^-$: the sparse estimate $\widehat{\bm{\Sigma}}^\ast_u$ is given by 
$\text{sign}(\bm{\Sigma}^-) \ \text{max}(|\bm{\Sigma}^-|- \lambda_\Sigma,0).$ If the minimum eigenvalue of the unconstrained solution is greater than 0, then the soft-thresholded matrix is the correct solution to \eqref{eq:Sigma obj func}. However, if the minimum eigenvalue of the soft-thresholded matrix is below 0, we follow \cite{bien2011sparse} and perform the optimization using the alternating direction method of multipliers (\citealp{boyd2011distributed}), which is implemented in the \proglang{R} function \texttt{ProxADMM} of the package \texttt{spcov} \citep{spcov}. 
Note that similarly to the estimation of the MF-VAR, we solve \eqref{eq:Sigma obj func} for a decrementing log-spaced grid of $\lambda_\Sigma$-values.
\end{remark}

If one wishes to incorporate the estimated nowcasting relations, the autoregressive parameters can be re-estimated by taking the error covariance matrix into account. 
This results in a type of  generalized least squares estimator of $\bm{\beta}$ as given by 
\begin{equation}\label{eq:GLS hierarchical estimator}
    \widehat{\bm{\beta}}^\ast = \underset{\bm{\beta}^\ast}{\text{argmin}} \ 
    \Big\{\frac{1}{2}\lVert \mathbf{y^\ast} - \mathbf{X^\ast} \bm{\beta}^\ast \rVert_2^2 + \lambda_{\beta^\ast} \mathcal{P}_{\text{Hier}}(\bm{\beta}^\ast) \Big \},
\end{equation}
where $\mathbf{y^\ast} = \widetilde{\bm{\Sigma}}^{-1/2} \mathbf{y}$,  $\mathbf{X^\ast} = \widetilde{\bm{\Sigma}}^{-1/2} \mathbf{X}$ and $\widetilde{\bm{\Sigma}} = \widehat{\bm{\Sigma}}^\ast_u \otimes \mathbf{I}_N$.

\section{Simulation Study}\label{Simulation Study}
We assess the performance of the proposed hierarchical group estimator through a simulation study where we compare its performance to three alternatives, namely the ordinary least squares (OLS), the ridge and the lasso. 

The set-up of the simulation study is driven by our empirical application (Section \ref{Macroeconomic Application}), namely the small MF-VAR with $K=22$ (one quarterly  and seven monthly variables) and $T=125$. The parameter matrix is set to reflect the obtained coefficients
which result in a stable MF-VAR. To make a clear distinction between zero and nonzero coefficients, we set all coefficients smaller than 0.01 to zero. As a result, the coefficient matrix does not strictly follow the recency-based hierarchical structure anymore, thereby favoring the hierarchical estimator less compared to its benchmarks. 
Throughout the paper, we standardized each  series to have sample mean zero and variance one \revision{as commonly done in the regularization literature before applying a sparse method such that all coefficients have comparable sizes after standardization}. To reduce the influence of initial conditions on the data generation process (DGP), we burn in the first 300 observations for each simulation run.

We consider \revision{four} 
simulation designs and run $R=500$ simulations in each. The first design compares the estimators in terms of their estimation accuracy and variable selection performance of the autoregressive parameter vector. In the second design, we analyze how well the proposed regularization method can detect the nowcasting relations between the low- and high-frequency variables in $\widehat{\bm{\Sigma}}_u^\ast$. The third design compares the point forecasts between the hierarchical estimator and its GLS-type version. \revision{The fourth design investigates the performance of the proposed estimator for DGPs with varying degree of sparsity.} 

\subsection{Autoregressive Estimation Accuracy and Variable Selection} \label{subsection Autoregressive Estimation Accuracy and Variable Selection}
We take the error covariance matrix to be the identity matrix and compare estimation accuracy of the autoregressive parameter vector by calculating the mean squared error
\begin{equation*}
    \text{MSE} = \frac{1}{R} \sum_{r=1}^R \frac{1}{K^2} \sum_{k=1}^{K^2} (\beta_k - \widehat{\beta}_k^{(r)})^2,
\end{equation*}
where $\widehat{\beta}_k^{(r)}$ refers to the $k$th element of the estimated parameter vector in simulation run $r$. 
To investigate variable selection performance, we use the false positive rate (FPR), the false negative rate (FNR) and Matthews correlation coefficient (MCC):
\begin{align*}
    &\text{FPR} = \frac{1}{R}\sum_{r=1}^R \frac{FP}{\# (k: \beta_k \neq 0)} \qquad \qquad
    \text{FNR} = \frac{1}{R}\sum_{r=1}^R \frac{FN}{\# (k: \beta_k = 0)} \\
    &\text{MCC} = \frac{1}{R} \sum_{r=1}^R \frac{TP \times TN - FP \times FN}{\sqrt{(TP+FP)(TP+FN)(TN+FP)(TN+FN)}},
\end{align*}
where $TP$ (and $TN$) are the number of regression coefficients that are estimated as nonzero (zero) and are also truly nonzero (zero) in the model and $FP$ (and $FN$) are the number of regression coefficients that are estimated as zero (nonzero), but are truly nonzero (zero) in the model.
Both FPR and FNR should be as small as possible. The MCC balances the two measures and is in essence a correlation coefficient between the true and estimated binary classifications. It returns a value between $-1$ and $+1$ with $+1$ representing a perfect prediction, $0$ no better than random prediction and $-1$ complete discrepancy between prediction and observation.
For the regularization methods (i.e., the proposed hierarchical estimator, lasso and ridge), we each time use a grid of 10 tuning parameters and select the one that minimizes the MSE between the estimated and true parameter vector.

\begin{figure}[t]
    \centering
    \makebox[\textwidth][c]{
    \includegraphics[scale = 0.5]{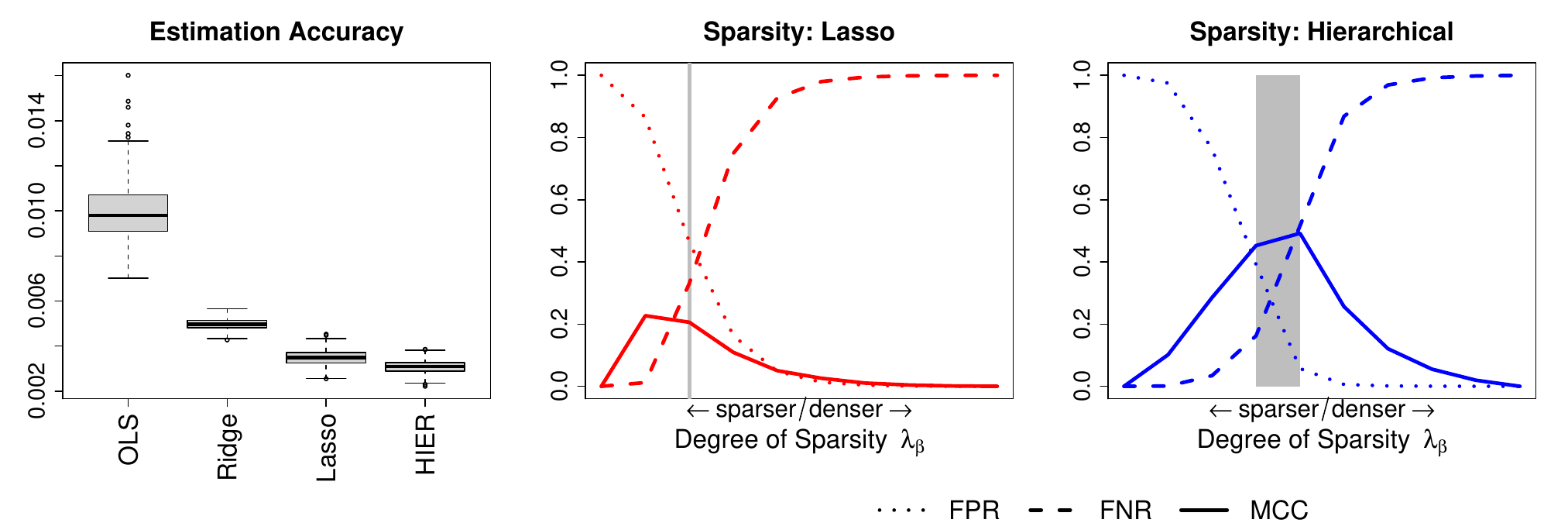}}
    \caption{Estimation accuracy for the four estimators (left) and variable selection performance of lasso and the hierarchical estimator (middle and right)}
    \label{fig:simulation_study_1}
\end{figure}

{\bf Results.} 
We first focus on estimation accuracy, see the left panel in Figure \ref{fig:simulation_study_1}. The hierarchical estimator generates the lowest estimation errors. It significantly outperforms all others in terms of MSE as confirmed by paired sample $t$-tests at 5\% significance level. OLS suffers as it is an unregularized estimator and thus cannot impose the necessary sparsity on the parameter vector; similarly for ridge which can only perform  shrinkage but not variable selection. The hierarchical estimator performs slightly better than lasso in terms of MSE, but the difference is less profound since lasso can also handle sparsity. 

Secondly, we compare the variable selection performance of the lasso and  hierarchical estimator, see the middle and right panel of Figure \ref{fig:simulation_study_1}, respectively. We plot the average FPR, FNR and MCC for different values of the tuning parameter $\lambda_\beta$. The variable selection performance of the hierarchical estimator is in line with its good performance in terms of estimation accuracy. The maximum MCC lies at roughly 0.5, in comparison to the maximum MCC of lasso which only reaches 0.23. The larger FPR of lasso indicates that its estimate is overly sparse, thereby missing important variables in the model. The larger FNR of the hierarchical estimator in comparison to lasso can be explained by the fact that the DGP does not favour the recency-based structure of the hierarchical estimator. Recall that we have set several small coefficients to zero. 
Thus, it is possible that within one parameter group we have large coefficients with lower priority and zero coefficients with higher priority. Alternatively, we can have hierarchical groups (all coefficients having the same priority) in which some coefficients are zero and some are large. 
To estimate and capture those important coefficients, our hierarchical estimator estimates some true zero coefficients as nonzero which automatically increases its FNR. Lastly, the grey area in the Figures indicates the 2.5\% and 97.5\% quantiles of the selected position in the tuning parameter grid across the simulation runs. It illustrates that the maximum MCC lies within the grey area of the hierarchical estimator but does not for lasso. 

\revision{
Finally, we repeat the simulation study using the hierarchical estimator with equal weights; see 
Figure 8 in Appendix B.1. While the equal weights version generates lower estimation errors,
the original weights version performs better 
in terms of sparsity recognition (MCC). 
 The hierarchical estimator with equal weights shrinks less aggressively, thereby having a higher FNR as many zero coefficients are estimated as small non-zeros.
}

\subsection{Nowcasting Relations}\label{Simulation Design: Nowcasting relations}
We evaluate the detection of the nowcasting relations.
To this end, we set the error covariance matrix to the regularized error covariance matrix estimated in the application Section \ref{Subsection Coincident Indicators}.
Coefficients smaller than 0.03 are set to zero to again ensure a clear distinction between the zeros and non-zeros.  
We first estimate the model using the four different estimators, calculate the resulting residual covariance matrix and then compute its regularized version through optimization problem \eqref{eq:Sigma obj func}. In line with our empirical application, we are mainly concerned with the sparsity pattern of the first row/column, namely the one corresponding to the low-frequency variable. Precisely, we investigate its variable selection performance using MCC, FPR and FNR. We estimate the MF-VAR and covariance matrix of the corresponding residuals for a two-dimensional (10 $\lambda_\beta$ $\times$ 10 $\lambda_\Sigma$) grid of tuning parameters. We select the tuning parameter couple that maximizes the MCC of the regularized covariance matrix in each simulation run.
Table \ref{tab:simulation study 2} contains the results.

\begin{table}[t]
    \centering
        \caption{Variable selection performance (MCC, FPR, FNR) of nowcasting relations in first row/column of the regularized error covariance matrix. ``Estimator" refers to the estimator used to estimate the MF-VAR$_{22}(1)$ and the corresponding residuals from which the covariance matrix is constructed. Standard errors are in parentheses.}
    \resizebox{0.9\textwidth}{!}{\begin{minipage}{\textwidth}        \centering
    \begin{tabular}{l c c c }
    \hline
        Estimator & MCC &  FPR & FNR \\ 
         \hline
        OLS  & 0.7111 (0.0052) &  0.1937 (0.0058)& 0.0960 (0.0064)\\            
        Ridge & 0.7568 (0.0050)  & 0.1637 (0.0051)& 0.0772 (0.0060)\\         
        Lasso & 0.7855 (0.0044)  &   0.1435 (0.0042)& 0.0655 (0.0055)\\       
        Hierarchical & 0.7856 (0.0045) &   0.1454 (0.0044)&  0.0640 (0.0055)\\  
        \hline
    \end{tabular}
\end{minipage}}
    \label{tab:simulation study 2}
\end{table}

{\bf Results.}
The performance across estimators is very similar, but the hierarchical estimator and lasso do perform best. Their MCCs lie at roughly 0.78. Their FPRs are slightly higher than their FNR which implies that the nowcasting relations tend to be estimated too sparsely. On the other hand, the low FNR suggests that in general we do not select variables which do not nowcast the low-frequency variable. 
Lastly, all estimators perform comparably across the tuning parameter grid for $\lambda_\Sigma$, but the variability around the selected $\lambda_\Sigma$ is higher for OLS and ridge than for lasso and the hierarchical estimator.

\subsection{Forecast Comparison} \label{Section: Simulation Study Forecast Comparison}
We assess whether  the best possible forecasting performance of the hierarchical estimator can be improved with its GLS version that incorporates the best nowcasting relations.

The error covariance matrix is set to the same matrix as in Section \ref{Simulation Design: Nowcasting relations}. We generate time series of length $T=105$ (as in the forecast  of Section \ref{Macroeconomic Application Sensitivity Analysis}), fit the models to the first $T-1$ observations and use the last observation to compute the one-step-ahead mean squared forecast error for each series. 
In line with the empirical application, we focus on forecast accuracy of the first series, 
which represents the low-frequency variable, 
and select the tuning parameter $\lambda_\beta$ that minimizes 
its squared forecast error.
For the selected model, we then estimate its regularized covariance matrix using 10 $\lambda_{\Sigma}$ values and choose the one that maximizes the MCC. 
Finally, with the selected $\widehat{\bm{\Sigma}}_u^\ast$, we re-estimate the model according to Equation \eqref{eq:GLS hierarchical estimator} to compare the forecast performance of the MF-VAR when the additional restrictions on the error covariance matrix are accounted for.

{\bf Results.}
The one-step ahead MSFE for the first series of the hierarchical unrestricted estimator and its GLS version are 0.5508 and 0.5810, respectively. 
The former significantly outperforms the latter,
as confirmed with a paired sample t-test at 1\% significance level. The addition of the nowcasting relation may not improve the forecast because the values of the covariances in the covariance matrix of the DGP, particularly of the first row/column, are relatively small in absolute terms.\footnote{The median absolute covariance in the first row/column lies at 0.0927 and the maximum
is 0.2046.} 
Moreover, it is important to point out that the running time for the estimation of the restricted version is substantially larger than for the unrestricted version. Thus, even if the covariance matrix would be denser, there is a clear trade-off between forecast accuracy and computational efficiency one needs to make. 

\subsection{\revision{Varying Degrees of Sparsity}}\label{subsection Varying Degree of Density}
\revision{In light of the recent debate on the validity of sparse methods for macroeconomic data (\citealp{giannone2021economic}), we 
revisit simulation studies 1 and 2 by varying the degree of sparsity of the DGP. 
Full details on the simulation designs 
are given in Appendix B.2.}

\revision{First, we  vary the degree of sparsity of the autoregressive parameters in simulation study 1.}
\revision{
Figure 9 demonstrates that the MSE of the hierarchical estimator stays relatively constant
when the degree of sparsity decreases.
Only for the fully dense DGP, we
observe a small increase in MSE. The hierarchical estimator with equal weights does not suffer from this.
Hence, if one expects a dense 
DGP 
or variable selection 
is not the main priority,  the hierarchical estimator with equal weight can be used. 
On the other hand, if one expects a sparse DGP or seeks parsimonious models to facilitate interpretation, more aggressive shrinkage can be enforced via the hierarchical estimator with proposed weights.}

\revision{Next, we vary the degree of sparsity in the first row/column of the error covariance matrix of simulation study 2. Here, denser settings}
\revision{
do have an effect on the detection of the nowcasting relations. Table 4 shows that 
the MCC decreases 
mainly due to an 
increase in the FPR, meaning that 
the nowcasting relations are estimated too sparsely. }

\section{Macroeconomic Application} \label{Macroeconomic Application}
We investigate a high-dimensional MF-VAR for the U.S.\ economy. We use data from 
1987 Q3 until 2018 Q4
$(T=126)$ on various aspects of the economy: amongst others output, income, prices and employment, see Table 5 in Appendix C.1 for an overview. The quarterly and monthly  series are directly taken from the FRED-QD and FRED-MD datasets which are available at the Federal Reserve Bank of St.\ Louis FRED database (see \citealp{mccracken2016fred}, \citeyear{mccracken2020fred} for more details). The weekly time series are additionally retrieved from the FRED database. The FRED-MD and -QD datasets contain transformation codes to make the data approximately stationary (see column ``T-code" in Table 5) which we apply to all series, \revision{thereby facilitating replicability and comparison with related research}.

To evaluate the influence of additional variables and  higher-frequency components, we estimate three MF-VAR models:\footnote{We label the three MF-VAR models as ``small", ``medium" and ``large" to compare their relative size. Note that even the small MF-VAR is large in traditional time series analysis.} 
The small MF-VAR ($K=22$) consists of the quarterly variable at interest, real GDP (\textit{GDPC1}), and seven monthly variables focusing on (industrial) production, employment and inflation. 
The medium MF-VAR ($K=55$) contains the small group and eleven additional monthly variables containing further information on different aspects of the economy including financial variables. 
The large MF-VAR incorporates the variables of the medium group and replaces four monthly variables (\textit{CLAIMSx, M2SL, FEDFUNDS, S\&P 500}) with their equivalent weekly series. See columns ``$K=22$", ``$K=55$" and ``$K=91$" of Table 5. 
To ensure that $m_2 = 12$ in the large MF-VAR, we consider all months with more than four weekly observations and disregard the excessive weeks at the beginning of the corresponding month, as in \cite{gotz2016testing}.

We aim to investigate several aspects. 
First, we focus on the autoregressive parameter estimates of the hierarchical estimator to investigate the predictive Granger causality relations between the series through a network analysis.
Second, we concentrate on nowcasting and investigate whether some high-frequency monthly and/or weekly economic series nowcast quarterly U.S.\ GDP growth and thus can deliver a coincident indicator of GDP growth. 
\revision{Third, we perform a sensitivity analysis of these results. Finally, we perform an out-of-sample expanding window nowcasting exercise including recent data on the Covid-19 pandemic.}
For ease of the discussion of the results, we follow the variable classification of \cite{mccracken2016fred} which can be found in Table 6 in Appendix C.1.

\subsection{Autoregressive Effects} \label{Subsection Variable Selection}
We first investigate predictive Granger causality relations.
To that end, we estimate the MF-VAR using the hierarchical estimator with proposed weights and recency-based priority structure (seasonally adjusted data are used) for all parameter groups and a grid of 10 tuning parameters $\lambda_{\beta}$. 
The tuning parameter is selected using rolling window time series cross-validation with window size $T_1$. 
For each rolling window whose in-sample period ends at time $t=T_1,\dots, T-1$, we first standardize each time series to have sample mean zero and variance one using the most recent $T_1$ observations, thereby taking possible time variations in the first and second moment of the data into account (see e.g., the Great Moderation \citealp{Campbell2007macroeconomic}; \citealp{stock2007why}).
Given the evaluation period $[T_1,T]$, we use the one-step-ahead mean squared forecast error as cross-validation score.
$T_1= 105$, leaving us with 20 observations for tuning parameter selection. 
We first discuss the results for the small MF-VAR, then summarize the findings for the medium and large MF-VARs.

\begin{figure}[t]
    \makebox[\textwidth][c]{
    \begin{subfigure}{0.5\textwidth}
        \includegraphics[scale = 0.37]{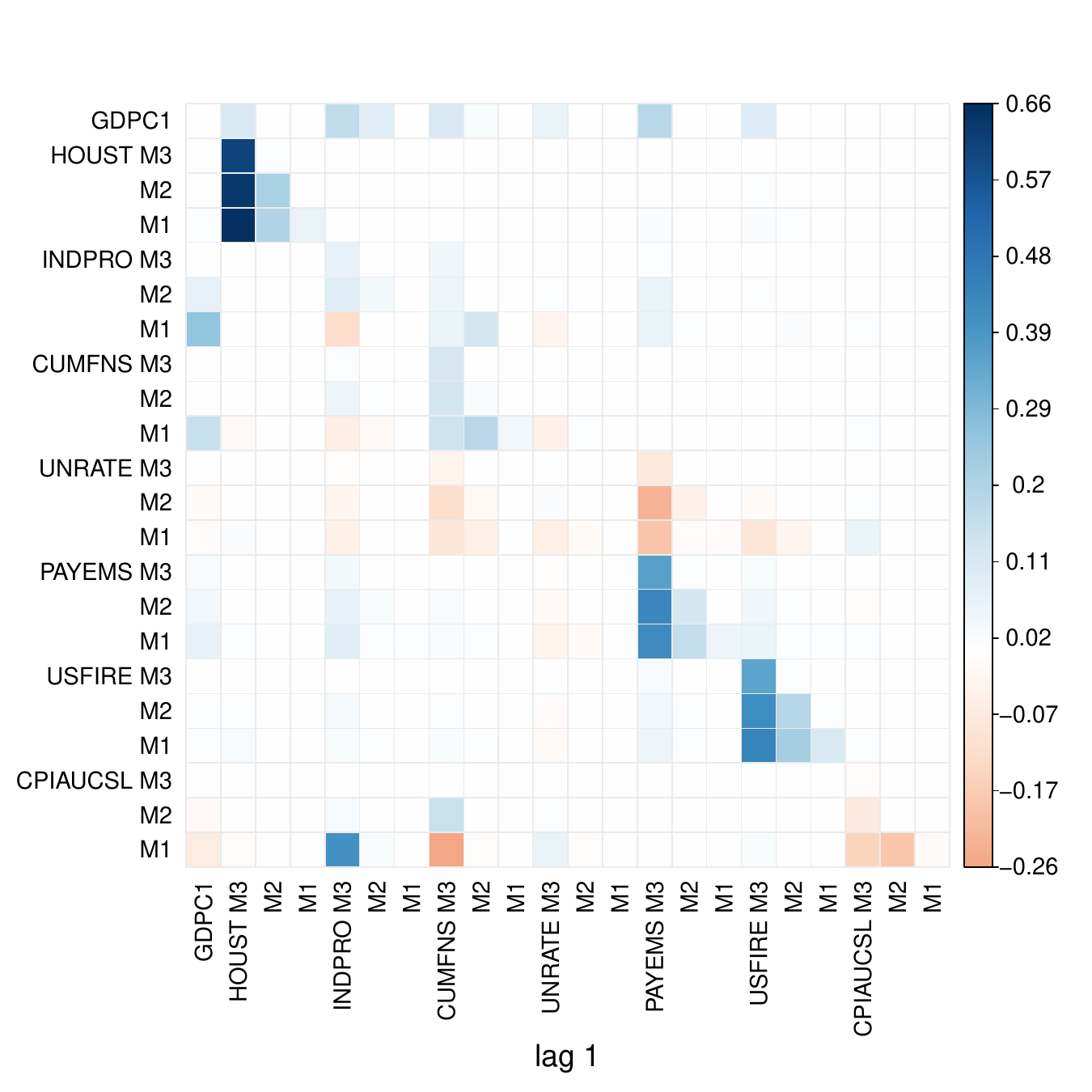}
        \caption{} 
    \end{subfigure}
    \begin{subfigure}{0.6\textwidth}
         \includegraphics[scale = 0.29]{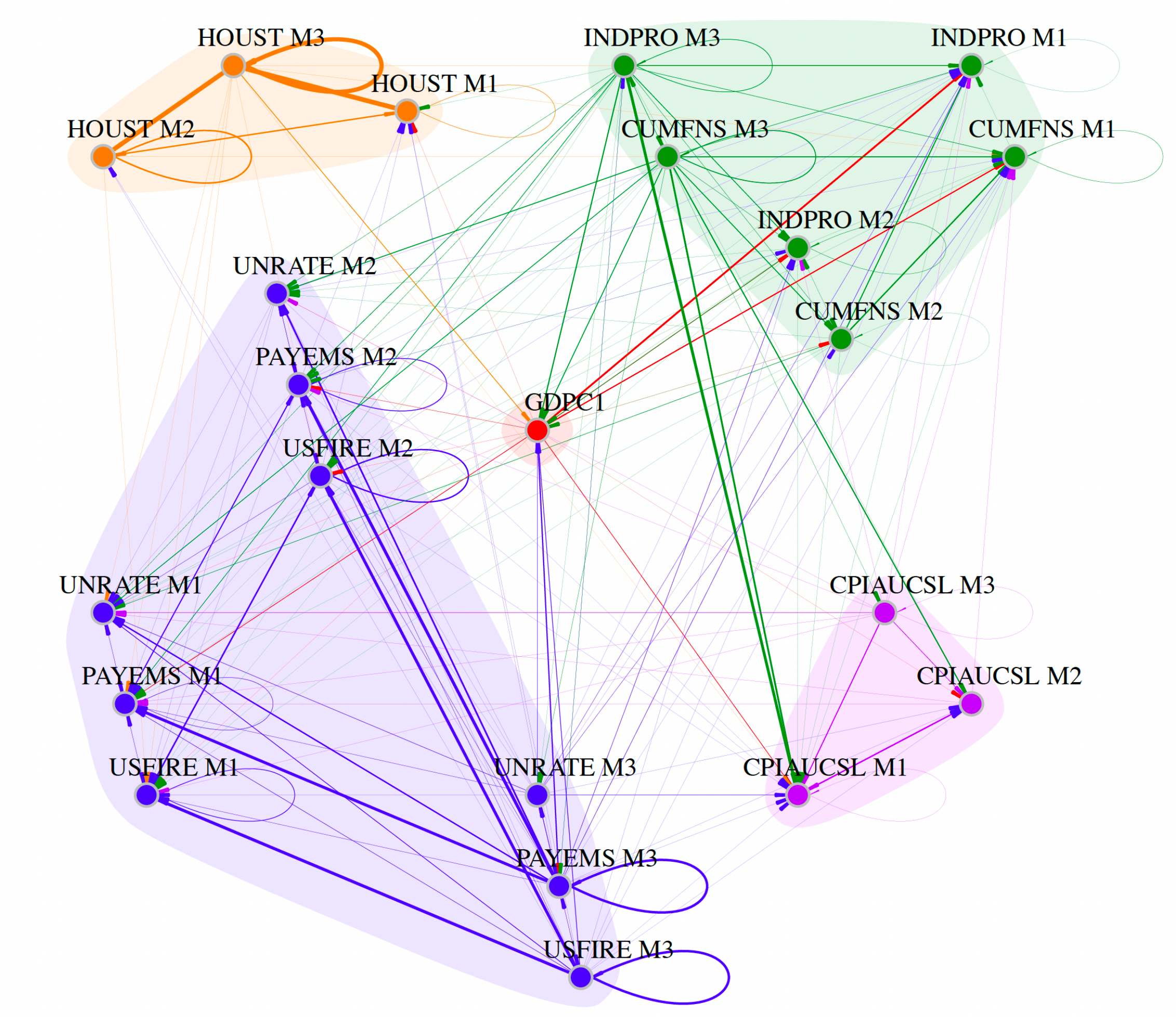}
        \caption{}
    \end{subfigure}
    }
    \caption{Small $(K=22)$ MF-VAR: 
    Autoregressive coefficients as a matrix (panel (a)) or as a directed network (panel (b)).
    The vertices represent the variables, the edges the nonzero coefficients. The edges' width are proportional to the absolute value of the estimates. Coloring of the vertices and their outgoing edges indicate the macroeconomic categories.}
    \label{fig:Variable Selection Small MFVAR}
\end{figure}

{\bf Small MF-VAR.}
Figure \ref{fig:Variable Selection Small MFVAR}(a) depicts the estimated autoregressive coefficient matrix of the hierarchical estimator; 197 of the 484 coefficients (roughly 41\%) are estimated as non-zero as indicated by the coloured cells. 
Figure \ref{fig:Variable Selection Small MFVAR}(b) visualizes the same results through a directed network. The vertices represent the variables, the edges the nonzero autoregressive coefficients. The edge's width is proportional to the absolute value of the estimate. The colors of the vertices and their outgoing edges indicate to which macroeconomic group in \cite{mccracken2016fred} the variable and its outgoing effect belong. 
We summarize the linkages between the macroeconomic categories in Table \ref{tab:Variable Selection small MFVAR per category}. The columns reflect a macroeconomic category's out-degree (influence), the rows its in-degree (responsiveness).

\begin{table}[t]
    \centering
    \caption{Small $(K=22)$ MF-VAR: Linkages between macroeconomic group. Entry $(i,j)$ indicates the number of edges from group $j$ to group $i$.}
    \resizebox{0.9\textwidth}{!}{\begin{minipage}{\textwidth}
    \centering
    \begin{tabular}{c|r r r r r |r }
    \hline
 To/From & GDP & Output \& Income & Housing	& Employment	& Prices &	 \textit{In-degree}\\
 \hline
GDP & 0 & 4 & 1 & 3 & 0 & 8 \\ 
  Output \& Income & 4 & 22 & 3  & 16 & 4 & 49\\ 
  Housing & 1 & 1 & 8 & 6 & 0 & 16\\ 
  Employment & 7 & 26 & 5 & 52 & 7 & 97\\ 
  Prices & 2 & 10 & 1  & 9 & 5 & 27\\ 
\hline
\textit{Out-degree}	& 14 & 63 & 18  & 86 & 16 & 197 \\
\hline
    \end{tabular}
    	\end{minipage} }
    \label{tab:Variable Selection small MFVAR per category}
\end{table}

We first concentrate on our main variable of interest \textit{GDPC1}. Eight variables contribute towards its prediction (see first row of Figure \ref{fig:Variable Selection Small MFVAR}(a) or incoming edges in panel (b)). 
Most influential is the macroeconomic group Output \& Income as month three and two are included for both variables (\textit{INDPRO} and \textit{CUMFNS}). The three variables related to employment (\textit{UNRATE, PAYEMS} and \textit{USFIRE}) and \textit{HOUST} are each selected with their third monthly component. 
Besides, 14 economic variables are influenced by lagged \textit{GDPC1} (see first column of Figure \ref{fig:Variable Selection Small MFVAR}(a) or outgoing edges of its vertex in panel (b)).
The Employment category has the most incoming edges, nevertheless, the most prominent (thicker) edges indicate that \textit{GDPC1} contributes the most to month one of \textit{INDPRO} and \textit{CUMFNS} (i.e., the Output \& Income group). 

Next, we focus on the linkages between macroeconomic groups containing the higher-frequency variables in Table \ref{tab:Variable Selection small MFVAR per category}. The categories Output \& Income and Employment are strongly connected with each other and have the most outgoing edges, even when taking into account that these groups contain more than one monthly variable. Output \& Income also contributes to the prediction of Prices and Employment towards Prices as well as Housing. While Prices play a substantial role for Output \& Income and Employment, Housing is more relevant for Employment than it is for Output \& Income. 

Finally, we inspect the linkages within each macroeconomic group (diagonal in Table \ref{tab:Variable Selection small MFVAR per category}). Employment and Housing display the highest within-group interaction. 
Zooming in on \textit{PAYEMS} or \textit{USFIRE} in Figure \ref{fig:Variable Selection Small MFVAR}, we  see that their own lagged effects are the strongest, despite them belonging to a group containing more than one high-frequency variable. Within  macroeconomic groups, a series' own lags thus remain the most informative. 

When focusing again on the network, we find that the third months of the variables have the most outgoing edges, whereas the first months have the most incoming edges. This is a logical consequence of the recency-based priority structure that we imposed. 

{\bf Medium and Large MF-VAR.}
To evaluate the influence of dimensionality, we now compare our results to the two larger MF-VARs.
Their networks are given in Figures 10 and 11 in Appendix C.2, whereas Tables 7 and 8 in Appendix C.1 present  the linkages between the macroeconomic categories, respectively.
The medium MF-VAR has 889 out of 3025 non-zero coefficients (30\%), the large MF-VAR 1199 out of 8281 (only 15\%). The increase in dimensionality thus induces a higher degree of selectiveness. 

Looking at the influencers of \textit{GDPC1},
we find that the majority of the variables selected for $K=22$ are also selected for the medium and large MF-VAR. 
In addition, the monetary (\textit{M2SL}) and financial variables (\textit{FEDFUNDS} and \textit{S\&P500})  as well as two monthly variables related to sales (\textit{CMRMTSPLx} and \textit{RETAILx}) deem to be relevant. 
Apart from month three of \textit{OILSPRICEx} and \textit{PCEPI} in the medium system, no variables related to prices are selected for the prediction of GDP growth. This coincides with the results for the small MF-VAR where \textit{CPIAUCSL} also only had minor influence. 
Similarly, the variables that \textit{GDPC1} influences in the medium and large system overlap with the selected ones in the small system. Particularly, \textit{GDPC1} strongly contributes to the prediction of the variables in the macroeconomic groups Output \& Income, illustrated by its thick outgoing edges, and Employment, indicated by the most incoming edges.

Next we focus on the linkages between the macroeconomic groups. For the medium MF-VAR, Table 7 underlines that Output \& Income, Sales, Employment and Prices are highly interconnected.
Moreover, the group Interest Rates influences the groups Employment and Prices. The latter one is not surprising as \textit{FEDFUNDS} is usually set to control inflation, hence one can expect changes in the previous quarter to aid in predicting inflation, measured by changes in prices. A similar argument supports the interconnection between Prices and Money. 
When looking at the diagonal entries of Table 7, we notice that similarly to the small MF-VAR, the groups Employment, Housing and Output \& Income have a high within-groups interaction. In contrast to the small system, the within-group linkages among Prices highly increases, which is likely due to the addition of four price variables.
The introduction of the weekly variables in the large system does not change the relations among the macroeconomic categories (see Table 8). 

\subsection{Coincident Indicators} \label{Subsection Coincident Indicators}
We investigate whether some high-frequency monthly and/or weekly economic series nowcast quarterly 
GDP growth and thus can deliver a coincident indicator. 
More specifically, we analyze how the performance of the indicator is influenced through (i) the formalization of ``sparse" nowcasting relations in the MF-VAR in comparison to a naive approach of constructing a coincident indicator based on the first principal component of \textit{all} high-frequency variables and (ii) the number of high-frequency components included in the model.

To construct the coincident indicator, we first estimate the MF-VAR using the hierarchical estimator as in Section \ref{Subsection Variable Selection}.
Secondly, we compute the regularized $\widehat{\bm{\Sigma}}_u^\ast$ from the MF-VAR residuals. We then select the high-frequency variables having a non-zero covariance element in the GDP column and construct the first principal component of the corresponding correlation matrix.\footnote{Alternatively, we constructed a coincident indicator by Partial Least Squares. Results are comparable.} We estimate the MF-VAR and corresponding covariance matrix of the MF-VAR residuals for a two-dimensional ($10 \times 10$) grid of tuning parameters $\lambda_\beta$ and $\lambda_\Sigma$. We report the results for the tuning parameter couple that maximizes the correlation between the most parsimonious coincident indicator and GDP growth.\footnote{We cannot select according to MCC so we use the 
maximal correlation as proxy.}

\begin{figure}[t]
\centering
    \begin{subfigure}[]{0.49\textwidth}
    \includegraphics[scale = 0.5]{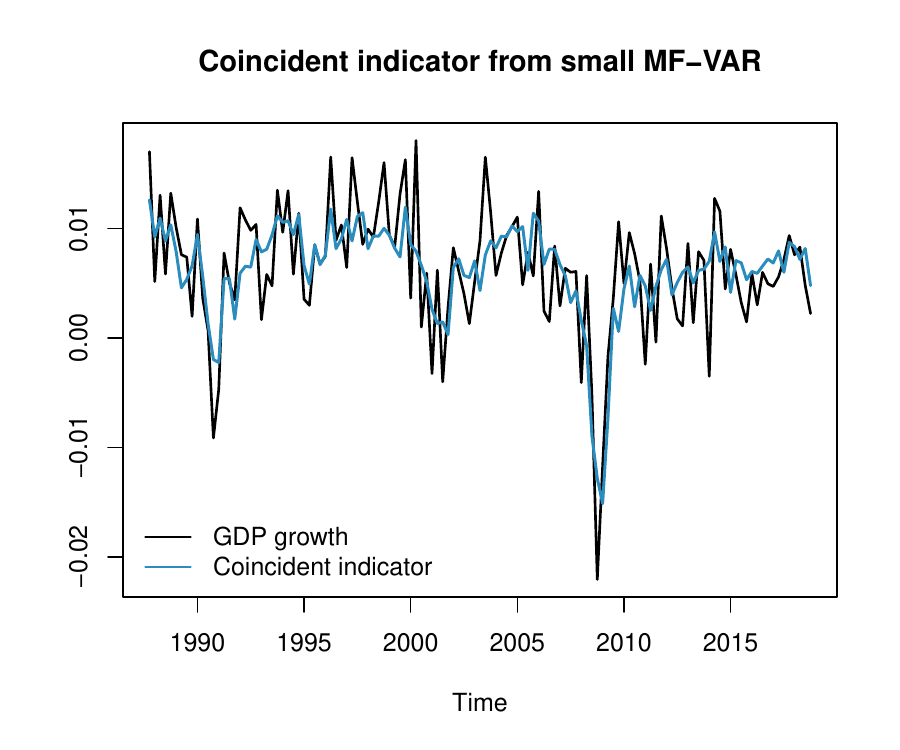}
    \caption{}
    \end{subfigure} 
    \begin{subfigure}[]{0.49\textwidth}
    \includegraphics[scale = 0.7]{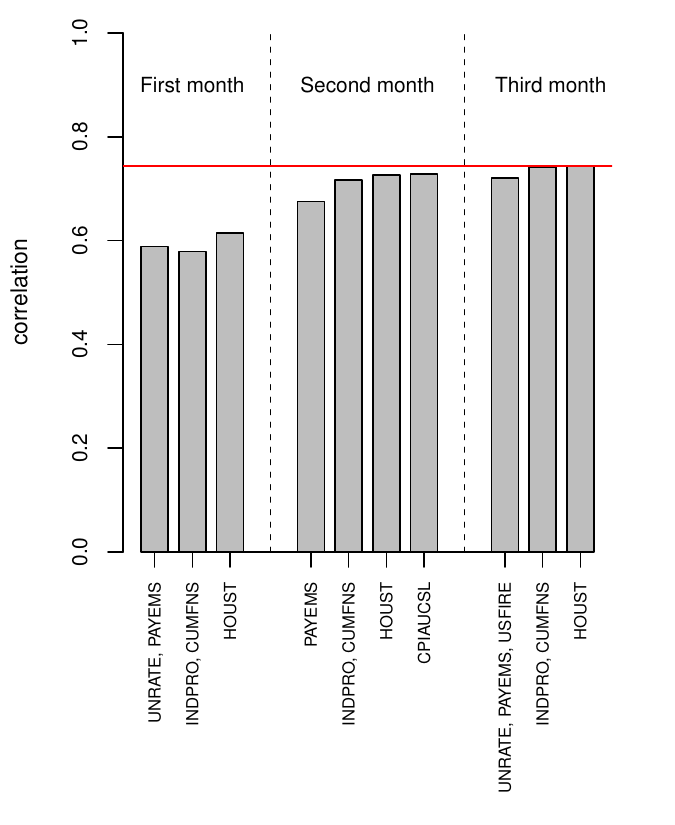}
    \vspace{-1.7cm}
    \caption{}
    \end{subfigure}  
    
    \caption{Small ($K=22$) MF-VAR: 
    Panel (a):  U.S.\ GDP growth versus coincident indicator. Panel (b):  Correlation between U.S.\ GDP growth and different coincident indicators by data release. The horizontal line indicates the correlation 
    with the coincident indicator constructed from all selected nowcasting relations from the entire quarter. }
    \label{fig:GDP vs indicator K = 22}
\end{figure}

{\bf Small MF-VAR.} 
Figure \ref{fig:GDP vs indicator K = 22}(a) plots the U.S.\ GDP growth against the coincident indicator for the small MF-VAR that maximizes the correlation to a value of  0.7429. The indicator tracks the movements of the GDP growth fairly well.
The selected monthly variables from which the first principal component has been constructed are listed on the x-axis of Figure \ref{fig:GDP vs indicator K = 22}(b). Thus, 16 out of the 21 high-frequency variables are selected, since their corresponding covariance with GDP growth is estimated as non-zero. Each variable is included with at least one monthly component. The variables related to Housing (\textit{HOUST}) and Output \& Income (\textit{INDPRO} and \textit{CUMFNS}) follow the fluctuations of GDP growth particularly well as all three months are selected. 
The variables measuring Employment (\textit{UNRATE}, \textit{PAYEMS} and \textit{USFIRE}) have varying levels of contribution. Lastly, the second month of \textit{CPIAUCSL}, which measures price changes, is also chosen.

For comparison, a coincident indicator constructed from \textit{all} high-frequency variables would result in a correlation of 0.7216. Capturing the nowcasting relations in a MF-VAR where the lagged and instantaneous dynamics are separated, thus increases the correlation by 2 percentage points. The advantage of carefully selecting variables before conducting a principal component analysis is a result consistent with \cite{bai2008forecasting}.
Furthermore, the correlations with GDP growth are fairly stable across the two-dimensional grid of tuning parameters.
In roughly half of the cases, the correlation is at least as high as the one computed from all high-frequency variables. Our findings are thus rather robust to a different choice of selection criterion for the tuning parameters.\footnote{Time series cross-validation (with one-step-ahead MSFE as CV-score) results in a similar correlation.}

\textit{GDPC1} is typically released with a relatively long delay (usually one month after the quarter for the first release), whereas the monthly variables are released in blocks at different dates throughout the following month.\footnote{In our case, the variables belonging to the same macroeconomic category (\citealp{mccracken2020fred}) are published on the same day of the month.} For instance, the previous month's \textit{UNRATE}, \textit{PAYEMS} and \textit{USFIRE} are typically released on the first Friday of the following month, whereas the remaining variables are only available around the middle of the month. We follow the release scheme of \cite{giannone2008nowcasting} to study the marginal impact of these data releases on the construction of our coincident indicator. 
We do not take data revisions into account. Hence, our vintages are ``pseudo" real-time vintages rather than true real-time vintages.

Figure \ref{fig:GDP vs indicator K = 22}(b) illustrates the achieved correlation between GDP growth and the coincident indicator, updated according to the release dates of the selected variables. As such, the first coincident indicator only uses month one for \textit{UNRATE} and \textit{PAYEMS}, whereas the last one is constructed from all selected variables and its correlation with GDP growth is indicated by the horizontal line. 
The figure shows that intra-quarter information matters. The second month releases have a large impact on the accuracy of the coincident indicator. Particularly, the addition of the variables \textit{INDPRO} and \textit{CUMFNS} raises the correlation significantly.
In fact, it is possible to construct an almost equally reliable indicator with the data from month one and two compared to one constructed with the data from the entire quarter. 
Hence, one can build a reliable coincident indicator roughly 1.5 months before the first release of GDP, thereby accounting for the publishing lag of approximately half a month for the monthly series.

{\bf Medium and Large MF-VAR.} 
Table \ref{tab:correlations overview} summarizes the correlations between GDP growth and the coincident indicators constructed from the different MF-VAR systems. The correlation for the large system (0.7928) is slightly higher than for the medium MF-VAR (0.7715) and both outperform the small system (0.7429).

\begin{table}[t]
    \centering
    \caption{Correlation between U.S.\ GDP growth and the coincident indicators for the small $(K= 22)$,  medium $(K= 55)$ and large $(K= 91)$ MF-VAR groups. }
        \resizebox{0.9\textwidth}{!}{\begin{minipage}{\textwidth}
    \centering
    \begin{tabular}{l c c c }
    \hline
        Type of coincident indicator& $K=22$ & $K=55$ & $K=91$ \\
        \hline
        Nowcasting relations M1& 0.6150 &  0.5128 &  0.4887 \\
        Nowcasting relations M1 + M2 & 0.7288 & 0.7495 & 0.6680\\
        All nowcasting relations & 0.7429 &0.7715 & 0.7928 \\
        All variables & 0.7216 & 0.7564 & 0.7557  \\
        \hline
    \end{tabular}
    \label{tab:correlations overview}
    \end{minipage}}
\end{table}

Full details on the medium and large MF-VAR are available in Appendix C.2: Figures 12(a) and 13(a) show that both indicators behave similarly and can better pick up the drop in GDP growth during the financial crisis than the coincident indicator of the small system. 
Many of the variables selected for the coincident indicator of the small MF-VAR are also selected for the two larger MF-VARs (see Figures 12(b) and 13(b)): There is a clear focus on variables related to (industrial) output, sales and employment whereas variables measuring price changes have a smaller influence. For the large MF-VAR 28 weekly variables are selected, thereby making it 
valuable to incorporate 
higher-frequency variables.
But the mere addition of variables does not lead to a larger correlation, emphasizing that selection is important:
Table \ref{tab:correlations overview} illustrates that the advantage of selecting the nowcasting relations in the MF-VARs persists as the correlation achieved from the first principal component with all variables is lower.
Lastly, the coincident indicator constructed from the selected variables from month one and two for the medium system performs comparable to the one constructed from all selected variables, in line with our finding for the small MF-VAR.

\subsection{\revision{Sensitivity Analyses}}\label{Macroeconomic Application Sensitivity Analysis}
\revision{
We investigate the sensitivity of our results 
regarding (i) the choice of weights for the hierarchical estimator,
(ii) the maximal lag length of the MF-VAR,
(iii) 
the inclusion of 
daily high-frequency 
data, and
(iv) forecast comparisons.
}

\revision{
{\bf Equal Weights.} 
We use the  hierarchical estimator with equal weights and present the autoregressive linkages
between the macroeconomic groups for the 
MF-VARs in
Tables 9-11 of Appendix C.3. 
Using equal weights corresponds to weaker penalization, hence 
the networks become denser, but the main conclusions regarding the inter-dependencies remain unchanged. Next, we revisit the results on the coincident indicators in
Table 12. 
For the small MF-VAR, the results are identical. For the medium and the large MF-VAR, 
respectively more and fewer high-frequency variables are selected for the coincident indicator which only affects its correlation with GDP growth in the first two months.}

\revision{
{\bf Lag Length.} We re-estimate the small MF-VAR with the hierarchical estimator using a maximal lag  $\ell = 1,2,$ or $4$.
Practically all coefficients 
from the second lag onwards (more than 98\%) are zeroed out, and  
more first-order coefficients are shrunken towards zero. 
Both the Bayesian and Akaike Information Criterion indicate one to be the ``optimal" lag length.}

 \revision{{\bf Daily Data.} To investigate the behavior of the hierarchical estimator in presence of daily high-frequency components, we replace the weekly financial variables (\textit{FEDFUNDS}, \textit{S\&P500})  in the large MF-VAR with 60 daily component series for each. This leads to an ``ultra-large" MF-VAR with $K=187$ component series instead of the large MF-VAR with $K=91$ series.
 Results are detailed in Appendix C.4.
 Only 
 5\% of the autoregressive coefficients are estimated as non-zero, 
 supporting our previous finding that an increase in dimensionality induces a higher degree of selectiveness. 
The influencers of \textit{GDPC1} stay roughly the same as in large MF-VAR. Interest Rate and Stock Price are selected with only two and three components respectively.
The coincident indicator
benefits from the usage of the more timely daily instead of weekly financial data only in the beginning of the quarter (for the first two months), but not when using the daily data from the whole quarter. 
}

{\bf Forecast Comparison.} 
We investigate the out-of-sample forecast performance of the hierarchical estimator, as detailed in Appendix C.5.
We compare its  performance to the random walk (RW) and AR(1) model as two popular,  simple univariate benchmarks; \revision{a traditional quarterly VAR(1), estimated by OLS, with all higher-frequency variables aggregated to the quarterly level;} and the three alternative estimators from Section \ref{Simulation Study} (OLS, ridge and lasso).
Table 15 provides the one-step ahead MSFE for \textit{GDPC1} 
across all MF-VARs. 

The hierarchical estimator generally outperforms the AR, RW, quarterly VAR and MF-VAR OLS, 
thereby indicating that higher-frequency variables as well as variable selection help to predict GDP growth. The large MF-VAR with  hierarchical estimator attains the lowest MSFE. 
Both the hierarchical estimator and lasso are included in the Model Confidence Set (\citealp{hansen2011model}) for two out of the three MF-VARs. 
\revision{The quarterly VAR is not included;  information contained in high-frequency series is thus beneficial in our application for forecasting GDP growth.}

\subsection{\revision{Nowcasting GDP growth pre and post Covid-19}}\label{subsection Nowcasting GDP pre and post Covid}
\revision{The Covid-19 pandemic has caused a dramatic drop in economic activity worldwide including the U.S. We end by assessing the performance of our coincident indicators in an out-of-sample expanding window nowcasting exercise with evaluation period 2019 Q1 until 2021 Q2, thereby covering the recent pandemic. For each current quarter in the expanding window approach, we estimate the MF-VAR with the hierarchical estimator (using data until the previous quarter), select the high-frequency variables for the
coincident indicator from the residual covariance matrix as in Section \ref{Subsection Coincident Indicators} and obtain their loadings on the first principal component. 
The nowcasts are then calculated by multiplying these loadings with the out-of-sample high-frequency data of the current quarter. 
} 

\begin{figure}[t]
    \begin{subtable}[]{0.49\linewidth}
    \resizebox{0.85\textwidth}{!}{\begin{minipage}{\textwidth}
    \begin{tabular}{l c c c}
    \hline
    &$K=22$ & $K=55$ & $K=91$ \\
    \hline
    Nowcasting & \multirow{2}{*}{0.1587} & \multirow{2}{*}{0.1385} & \multirow{2}{*}{0.1782}\\
    relations M1 & \\
    Nowcasting & \multirow{2}{*}{0.0661} & \multirow{2}{*}{0.0340} & \multirow{2}{*}{0.0431}\\
    relations M1 + M2  & \\
    All nowcasting  & \multirow{2}{*}{0.0337} & \multirow{2}{*}{0.0875} & \multirow{2}{*}{0.0919}\\
    relations \\
    \hline
    \end{tabular}    \end{minipage}}
    \caption{} 
    \end{subtable}
        \begin{subfigure}[]{0.49\textwidth}
    \includegraphics[scale = 0.5]{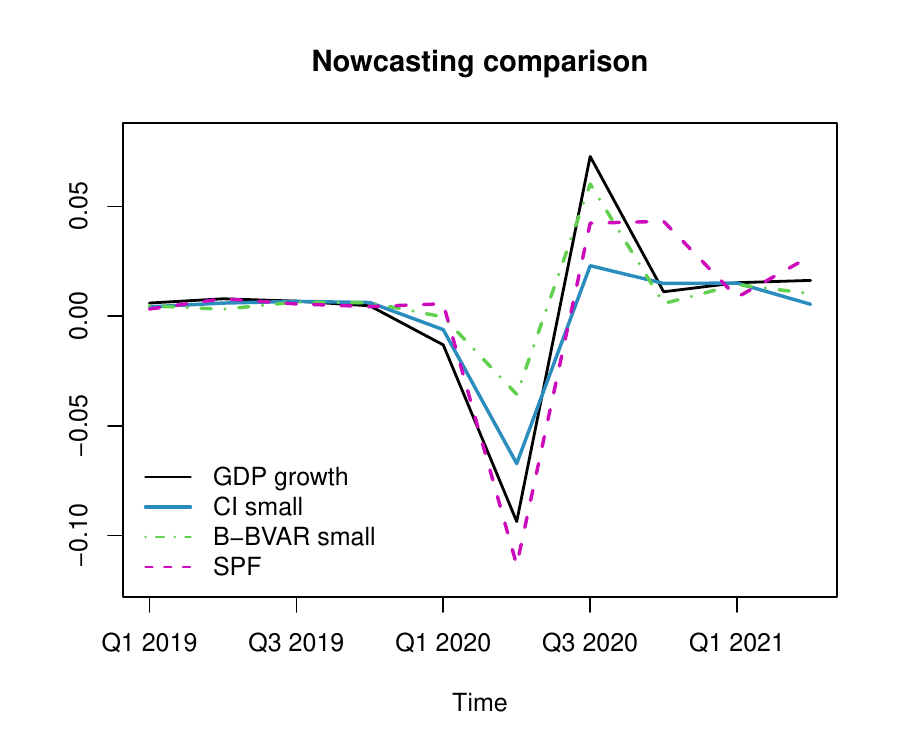}
    \caption{}
    \end{subfigure} 
    \caption{\revision{Panel (a): Nowcasting performance (mean squared nowcast errors of GDP growth, multiplied by 100) of the three MF-VARs for 2019 Q1 to 2021 Q2. Panel (b): U.S.\ GDP growth versus our coincident indicator (CI), and the B-BVAR and SPF nowcasts.}}
    \label{fig:nowcasting performance tab and fig}
\end{figure}

\revision{
The mean squared nowcast errors between GDP growth and the coincident indicators constructed from the three MF-VARs are depicted on the last line of the table in Figure \ref{fig:nowcasting performance tab and fig}(a).
The coincident indicator of the small MF-VAR attains the best accuracy, 
the error more than doubles for the larger MF-VARs.
In addition, we track the nowcasting accuracy as the quarter progresses, thereby constructing the coincident indicator only from the selected high-frequency variables from the first month (first line), or the first two months (second line in the table).
For the best performing small MF-VAR, the nowcasts gradually improve as new data gets released along the quarter, with the largest improvement occurring when adding high-frequency data from the second month to the first.
}

\begin{figure}[t]
 \makebox[\textwidth][c]{
    \begin{subfigure}{0.5\linewidth}
\includegraphics[scale = 0.45]{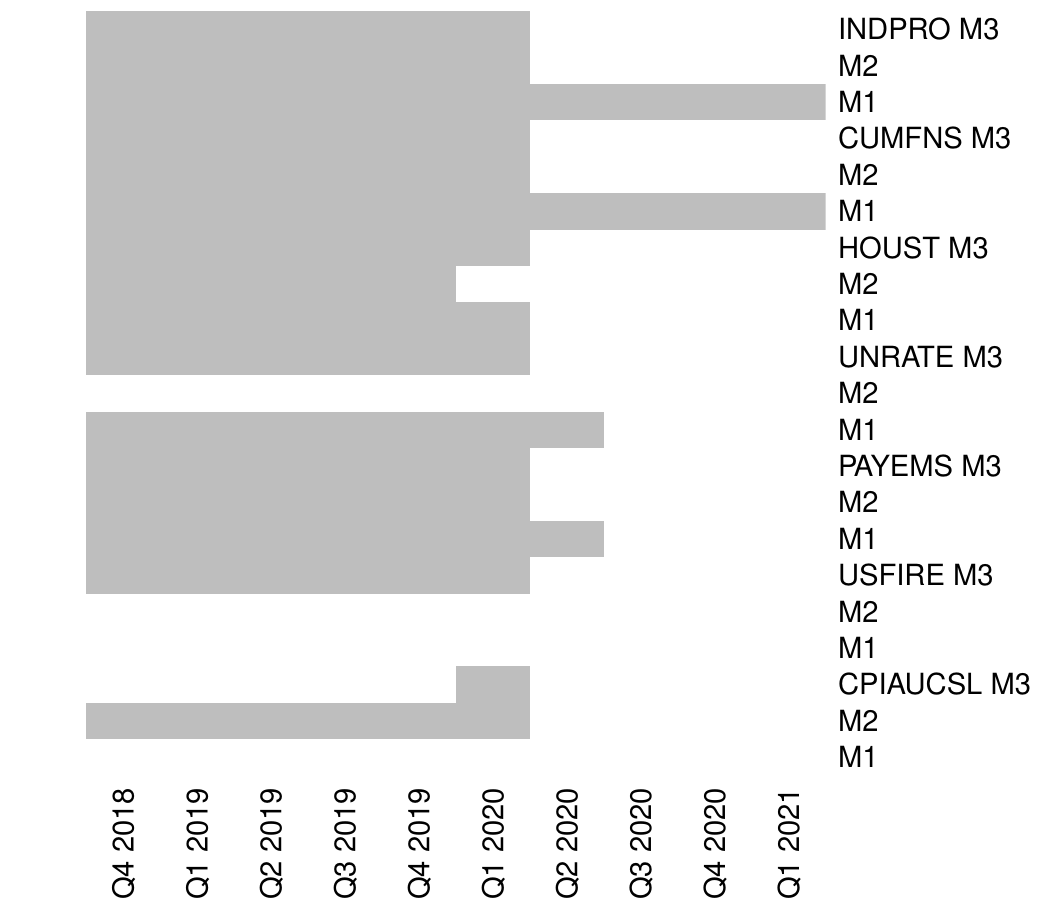}
    \end{subfigure}
    \begin{subfigure}{0.5\linewidth}
     \includegraphics[scale = 0.45]{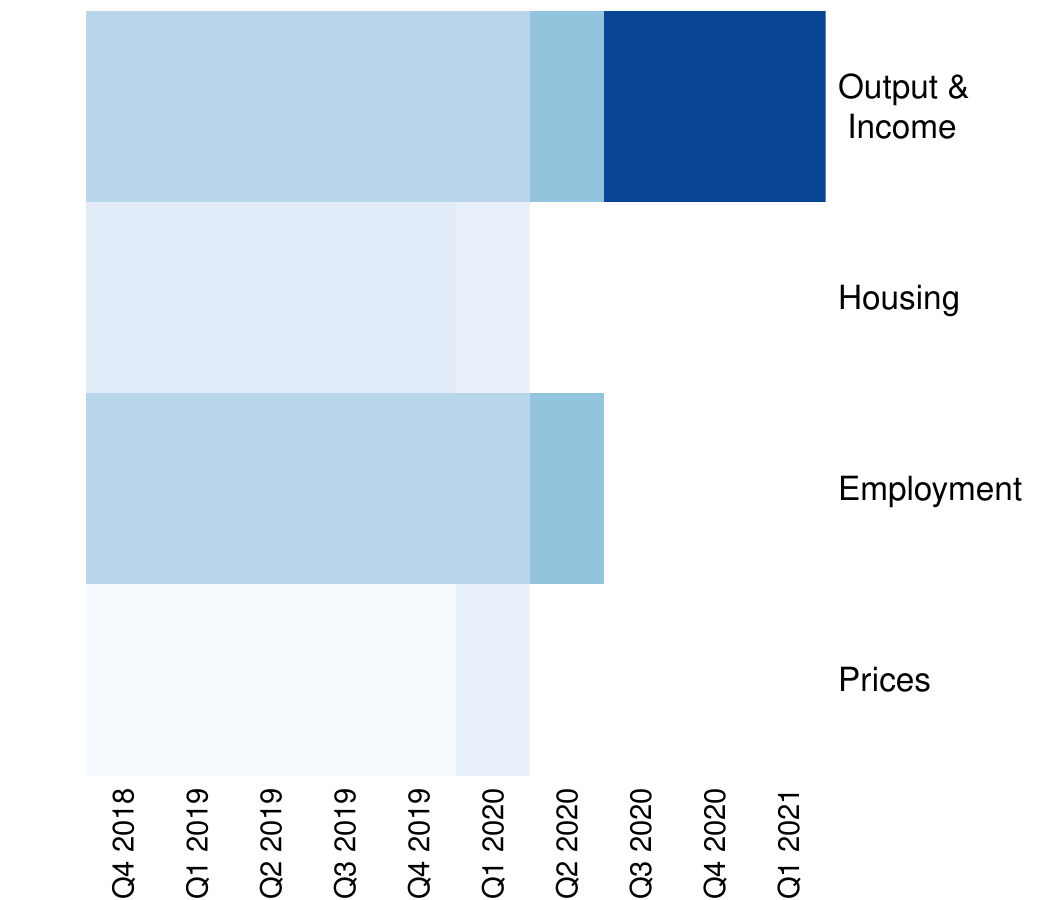}
    \end{subfigure}}
    \caption{\revision{Small ($K=22$) MF-VAR: Panel(a): Selected high-frequency variables for the coincident indicator. Panel(b): Fraction of non-zero coefficients in each macroeconomic category to the total number of non-zero coefficients in the coincident indicator. The horizontal axis represents the ending date of each expanding window.}}
    \label{fig:nowcasting development selection}
\end{figure}

\revision{Next, we discuss the stability through time of the selected variables from which the coincident indicator of the  small MF-VAR has been constructed. Panel(a) of Figure \ref{fig:nowcasting development selection} shows the selected high-frequency variables for each end point of the expanding window. Panel(b) summarizes them according to the macroeconomic group they belong to. Before the pandemic hits the U.S.\ economy in 2020 Q1, the selection of the high-frequency variables was very stable. Almost all variables are selected with all monthly components, the exceptions are \textit{UNRATE}, \textit{USFIRE} and \textit{CPIAUCSL}.
At the onset of the pandemic, a clear break in variable selection occurs: the coincident indicator is much more sparsely constructed thereby only using data on the Output \& Income group from the first month.} 

\revision{Finally, we compare the performance of the coincident indicator of the small MF-VAR with two state-of-the-art benchmarks, one Bayesian and one survey. We use the ``blocking" Bayesian MF-VAR (B-BVAR) by \cite{cimadomo2021nowcasting} since it also relies on the stacked MF-VAR approach.}\footnote{\revision{We would like to thank Michele Lenza for providing the code for the B-BVAR method which we applied \revision{(with default settings for stationary data) to the small and  medium MF-VAR. 
In line with our findings, their small MF-VAR was best performing, hence the results for their medium MF-VAR are omitted but available upon request.}}}
\revision{Second, we use the GDP nowcasts of the Survey of Professional} \revision{Forecasters (SPF). Figure \ref{fig:nowcasting performance tab and fig}(b) plots  GDP growth against the three different nowcasts for 2019 Q1 to 2020 Q2.\footnote{\revision{Please note that all insights obtained from Figure \ref{fig:nowcasting performance tab and fig}(b) are still subject to GDP being revised.}} Overall, all methods track the same fluctuations in the business cycle.
Before the pandemic, all nowcasts are practically the same and very close to actual GDP growth. 
When the pandemic hits the economy, some differences can be observed. 
First, the SPF does not pick up the slight drop in GDP growth in Q1 2020. This can be explained by the fact that the SPF nowcasts are released during the second month of a quarter, while most countries only went into lockdown by March. 
Second, the drop in economic activity in 2020 Q2 is very pronounced for the SPF, much less so for our coincident indicator and the B-BVAR nowcast.
The recovery in Q3 thereafter is best tracked by the B-BVAR. Our coincident indicator and the B-BVAR nowcasts align again from Q4 onwards.}

\section{Conclusion} \label{Conclusion}
We introduce a convex regularization method tailored towards the hierarchically ordered structure of mixed-frequency VARs. 
To this end, we use a group lasso with nested groups which permits various forms of hierarchical sparsity patterns that allows one to discriminate between recent and obsolete information. Our simulation study shows that the proposed regularizer can improve estimation and variable selection performance. 
Furthermore, nowcasting relations can be detected from the sparsity pattern of the covariance matrix of the MF-VAR errors. Those high-frequency variables that nowcast the low-frequency variables, as evident from their non-zero contemporaneous link, can deliver a coincident indicator of the low-frequency variable. Constructing coincident indicators from a group of selected variables rather than all permits policy makers to get an earlier grasp of the state of the economy, as can be seen from our economic application on U.S.\ GDP growth. 

\revision{
The proposed MF-VAR method is quite flexible and can be extended in various ways. 
First, regularization via restrictions other than sparsity can be explored. Temporal aggregation restrictions, for instance, can be imposed in the MF-VAR by exploiting fusion \revision{penalties (e.g., \citealp{yan2021rare}) that encourage similarity across certain coefficients.}}
\revision{For monthly stock data it could, for instance, be interesting to encourage the effects of all months  on the quarterly variable to be similar, thereby implicitly aggregating the monthly variable to the quarterly level. 
Second, an interesting path for future research concerns the extension of our method to enable structural analysis 
as recently done in a Bayesian set-up by e.g., \cite{cimadomo2021nowcasting} who use generalized impulse responses to track transmission mechanisms of low-frequency shocks hitting the U.S.\ economy, 
or \cite{paccagnini2021identifying} who use
orthogonalized impulse responses to identify the impact of high-frequency shocks thereby revealing a temporal aggregation bias when adopting single low-frequency models instead of mixed-frequency ones.
Lastly, while we consider MF-VAR for stationary data,
a natural next step would be to 
allow for non-stationarity 
\revision{by building on the lag-augmentation idea of \cite{toda1995statistical}
as done in  \cite{gotz2019grangernonstationary} for low-dimensional mixed MF-VARs or \cite{hecq2021inference} for high-dimensional VARs.
}}

\section*{Supplemental Materials}
\begin{description}
\item[\texttt{R}-code:] 
Supplemental files for this article include \texttt{R}-code  to reproduce all results.
Please consult the file
README in the zip file for more details. (Code.zip, zip archive)
\item[Appendix:] 
The Appendix contains 
implementation details on the algorithm, and additional results on the simulations and empirical application.  (Appendix.pdf)
\end{description}

\section*{Acknowledgments} 
We thank the editor, associate editor and reviewers for their thorough review and highly appreciate their constructive comments which substantially improved the quality of the paper. Wilms gratefully acknowledges funding from  the European Union's Horizon 2020 research and innovation programme (Marie Sk\l{}odowska-Curie grant  No 832671).

\bibliographystyle{Chicago} 
\bibliography{references}

\newpage
\renewcommand*\appendixpagename{\Large\bfseries Appendices}
\appendixpage
\appendix
\section{Algorithm}\label{Appendix Algorithms}

Algorithm \ref{algorithm1} provides an overview of the proximal gradient algorithm (see e.g., \citealp{tseng2008accelerated}) to efficiently solve the optimization problem in 
Equation (\ref{eq:optimization problem}). 
A key ingredient of the algorithm concerns the proximal operator $\text{Prox}_{v \lambda \mathcal{P}(\cdot)}$ which has a closed-form solution making it extremely efficient to compute. Indeed, 
the updates of each hierarchical group $p=1,\dots, P_g$ correspond to a groupwise soft thresholding operation given by $\text{max}(1-v\lambda \cdot w_{s_g^{(p)}}/\lVert \widetilde{\bm{\beta}}_{s_g^{(p)}}\rVert_2, 0) \ \widetilde{\bm{\beta}}_{s_g^{(p)}}$ with $v$ being the step size which we set equal to the largest singular value of $\mathbf{X}$.
Algorithm \ref{algorithm1} requires a starting value $\bm{\beta}[0]$ which we initially set equal to $\bf{0}$. Finally, while the algorithm 
is given for a fixed value of the tuning parameter $\lambda$, it is standard in the regularization literature to implement it for a decrementing log-spaced grid of $\lambda$ values.
The starting value $\lambda_{\text{max}}$ is an estimate of the smallest value that sets all coefficients equal to zero. For each smaller $\lambda$ along the grid, we use the outcome of the previous run as a warm-start for $\bm{\beta}[0]$.

\begin{algorithm}
	\caption{Accelerated proximal gradient method} \label{algorithm1}
	\begin{algorithmic} 
		\REQUIRE $\mathbf{y}$, $\mathbf{X}$, $\mathcal{P}(\bm{\beta})$, $\lambda$, $\varepsilon$ \\
		initialization: \\
		\begin{itemize}
			\itemsep0em 
			\item $\bm{\beta}[1] \leftarrow \bm{\beta}[2] \leftarrow  \bm{\beta}[0]$
			\item step size $v$ which is set equal to the largest singular value of $\mathbf{X}$
		\end{itemize}
		
		\FOR {$r = 3,4,...$} 
		\FOR {$g = 1,...,G$}
		
		\STATE $\ddot{\bm{\beta}}_{g} \leftarrow \bm{\beta}_{g}[r-1] + \frac{r-2}{r+1} \Big(\bm{\beta}_{g}[r-1]-\bm{\beta}_{g}[r-2] \Big)$ 
		\STATE $\bm{\beta}_{g}[r] \leftarrow\text{Prox}_{v \lambda \mathcal{P}(\cdot)} \Big( \ddot{\bm{\beta}}_g - v \nabla_{\bm{\beta}_{g}} \mathcal{L}(\ddot{\bm{\beta}}_g)\Big)$
		\STATE \text{where} $\nabla_{\bm{\beta}_{g}} \mathcal{L}(\ddot{\bm{\beta}}_g) = - (\mathbf{y} - \mathbf{X}_{\bm{\beta}_{g}} \ddot{\bm{\beta}}_g - \mathbf{X}_{\bm{\beta}_{f>g}} \bm{\beta}_{f>g}[r-1] - \mathbf{X}_{\bm{\beta}_{f<g}} \bm{\beta}_{f<g}[r])^{\prime} \mathbf{X}_{\bm{\beta}_{g}}$
		\ENDFOR
		\IF {$\lVert \bm{\beta}[r] - \bm{\beta}[r-1] \rVert_{\infty} \leq
			\varepsilon$}
		\STATE $\textbf{break}$
		\ENDIF
		
		\ENDFOR
		\RETURN $\bm{\beta}[r]$
	\end{algorithmic}
\end{algorithm}

\section{Simulation Study}\label{Appendix Simulation Study}
\subsection{Sensitivity Analysis: Choice of weights for hierarchical estimator}\label{Appendix Sensitivity Analysis Choice of weights}

\begin{figure}[H] 
	\centering
	\makebox[\textwidth][c]{
		\includegraphics[scale = 0.5]{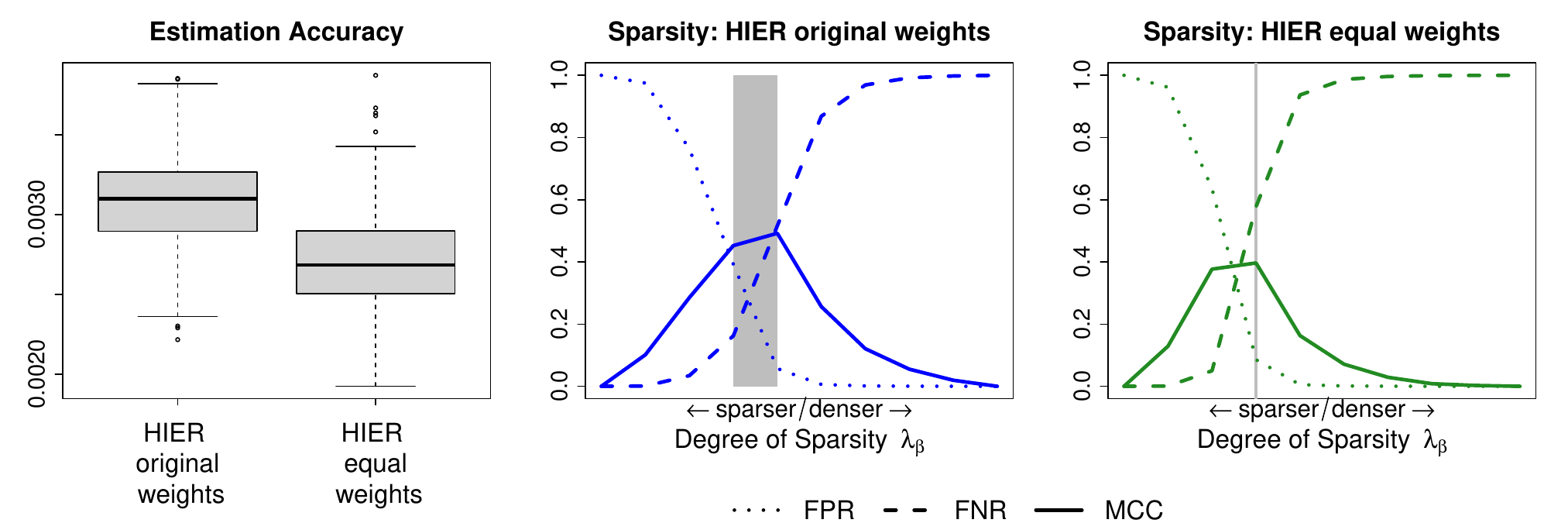}}
	\caption{Estimation accuracy for the hierarchical estimator with original and equal weights (left) and variable selection performance with original (middle) and equal weights (right).}
	\label{fig:simulation_study_1_weights}
\end{figure}

Figure \ref{fig:simulation_study_1_weights} compares the results of simulation study 1 for the hierarchical estimator with the proposed weights $w_{s_g^{(p)}}  = \text{card}(s_g^{(1)})-\text{card}(s_g^{(p)})+1$ and equal weights. 

\subsection{Sensitivity Analysis: Denser DGPs}\label{Appendix Sensitvity Analysis sparse and dense DGPs}
To assess the effect on the hierarchical estimator of varying degrees of sparsity in the autoregressive parameters and the covariance matrix of the error terms, we repeat simulation study 1 and simulation study 2 with varying degrees of sparsity in their respective DGPs. 

\begin{figure}
	\centering
	\includegraphics[scale = 0.5]{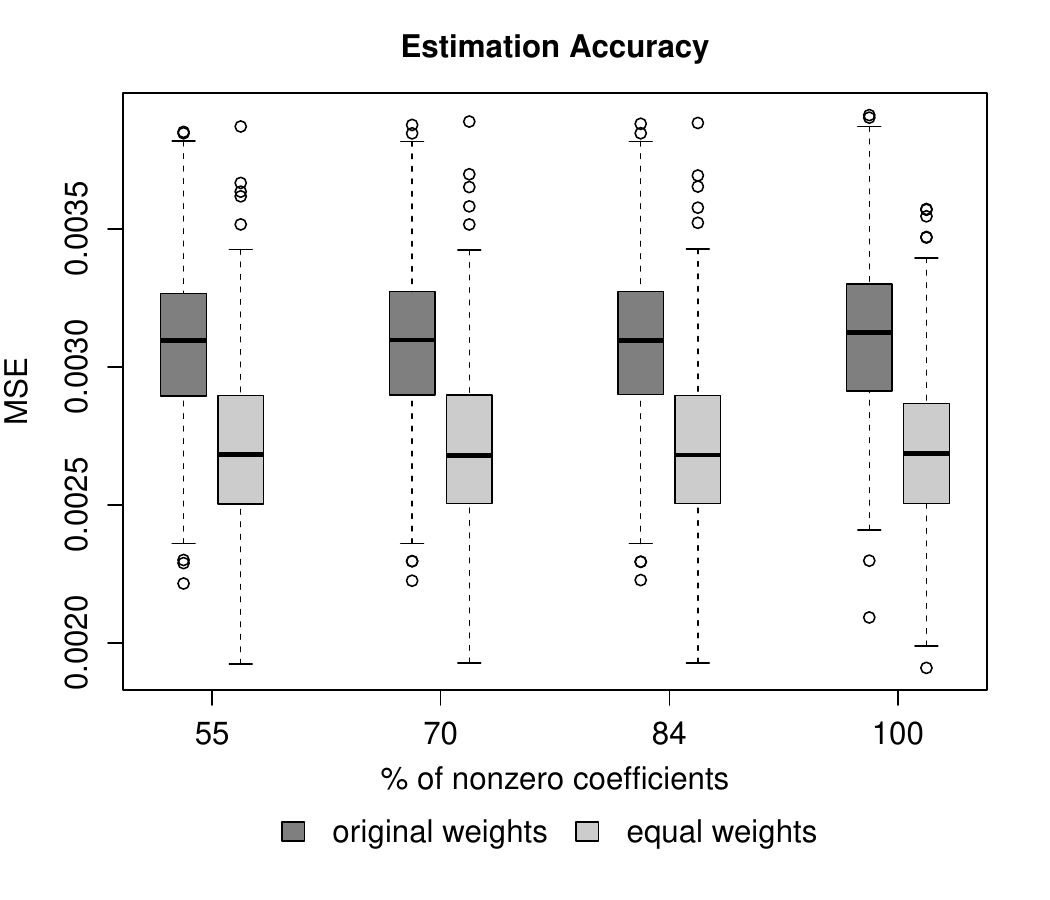}
	\caption{Estimation accuracy for the hierarchical estimator with original (dark grey) and equal weights (light grey) across DGPs with varying degrees of sparsity in the AR coefficients.}
	\label{fig:simulation_sparse_dense_weights}
\end{figure}

For simulation study 1, we gradually increase the percentage of non-zero coefficients in the autoregressive parameters 
across four settings. The sparsest setting (55\% of non-zero coefficients) starts from the design used in Section \ref{subsection Autoregressive Estimation Accuracy and Variable Selection} where all coefficients smaller than 0.01 are set to zero.
For the second setting (70\% non-zeros), all coefficients smaller than 0.002 are set to zero. 
For the third setting (84\% non-zeros), we take the unaltered parameter matrix from Section \ref{subsection Autoregressive Estimation Accuracy and Variable Selection}.
For the densest setting (100\% non-zeros) all zero coefficients of the sparsest setting are assigned a value of 0.01.
Figure \ref{fig:simulation_sparse_dense_weights} gives the results.

For simulation study 2, we gradually increase the percentage of non-zero coefficients in the first row/column of the error covariance matrix across four settings. The sparsest setting corresponds to the design of Section \ref{Simulation Design: Nowcasting relations} with 62\% non-zeros. We then increase the density towards respectively 76\%, 90\% and 100\% non-zeros by each time setting the remaining zero coefficients in the first row/column to 0.03. Table \ref{tab:simulation_sparse_dense_varcov} displays the results.

\begin{table}[h]
	\centering
	\caption{
		Variable selection performance (MCC, FPR, FNR) of nowcasting relations for varying degrees of sparsity in first row/column of the regularized error covariance matrix. Standard errors are in parentheses.}
	\begin{threeparttable}
		\begin{tabular}{lcccc}
			\hline
			& \multicolumn{4}{c}{\% of non-zero coefficients in first row/column}\\
			& 62 & 76 & 90 & 100 \\
			\hline
			MCC & 0.7792 (0.0044)&  0.7037 (0.0046)&  0.5427 (0.0093) & NA\tnote{1} \\
			FPR & 0.1448 (0.0045)& 0.1301 (0.0057) & 0.2040 (0.0066) &  0.2352 (0.0081)\tnote{2} \\
			FNR & 0.0735  (0.0059)& 0.1488 (0.0059) &  0.0400 (0.0067)  & NA\tnote{1} \\
			\hline
		\end{tabular}
		\begin{tablenotes}\footnotesize
			\item[1] FNR and MCC are not defined since number of TN is zero.
			
			\item[2] Tuning parameter selection based on maximizing the correlation between the most parsimonious coincident indicator and the first series since the MCC is not available. 
		\end{tablenotes}
	\end{threeparttable}
	\label{tab:simulation_sparse_dense_varcov}
\end{table}

\section{Macroeconomic Application} \label{Appendix Application}
\subsection{Additional Tables} \label{Appendix Application Tables}

\begin{table}[H]
	\centering
	\caption{Data description table. Column T-code denotes the data transformation applied to a time-series, which are: (1) not transformed, (2) $\Delta x_t$, (3) $ \Delta^2 x_t$, (4) $\text{log}(x_t)$, (5) $\Delta \text{log}(x_t)$, (6) $\Delta^2 \text{log}(x_t)$.
		Columns $K=22$, $K=55$ and $K=91$ indicate whether and at which frequency the variable was included in the model.}
	\resizebox{0.9\textwidth}{!}{\begin{minipage}{\textwidth}
			\centering
			\begin{tabular}{llcccc}
				\hline
				FRED Code & Description &T-code & $K=22$ & $K=55$ & $K=91$ \\
				\hline
				GDPC1 & Real Gross Domestic Product & 5 & Q & Q & Q\\
				RPI & Real Personal Income & 5 & & M & M \\
				CMRMTSPLx & Real Manufacturing and & 5 & & M & M \\ & Trade Industries Sales \\
				RETAILx & Retail and Food Services Sales & 5 & & M & M \\	
				HOUST & Housing Starts: Total & 4 & M & M & M \\	
				INDPRO & IP Index & 5 & M & M & M \\
				CUMFNS & Capacity Utilization: Manufacturing & 2 & M & M & M  \\
				UNRATE & Civilian Unemployment Rate & 2 & M & M & M \\
				PAYEMS &All Employees: Total nonfarm & 5 & M & M & M \\
				USFIRE & All Employees: Financial Activities  & 5 & M & M & M \\
				CLAIMSx & Initial Claims  & 5 &  & M & W\\
				CPIAUCSL & CPI : All Items  & 6 & M & M & M\\
				CPIULFSL & CPI : All Items Less Food  & 6 &  & M & M\\
				PCEPI & Personal Cons. Expend.: Chain Index  & 6 & & M & M \\
				WPSFD49207 & PPI: Finished Goods & 6 &  & M & M\\
				OILPRICEx & Crude Oil  & 6 &  & M & M\\
				M2SL & M2 Money Stock  & 6 &  & M & W\\	
				FEDFUNDS & Effective Federal Funds Rate & 2 &  & M & W\\
				S\&P 500 & S\&P’s Stock Price Index  & 5 &  & M & W\\ 
				\hline
			\end{tabular}
	\end{minipage}}
	\label{tab:data_description}
\end{table}

\begin{table}
	\centering
	\caption{Categories of monthly series in the medium ($K=55$) MF-VAR group, following the grouping of \cite{mccracken2016fred} in their Appendix.}
	\resizebox{0.9\textwidth}{!}{\begin{minipage}{\textwidth}
			\centering
			\begin{tabular}{l | l}
				\hline
				Macroeconomic category & FRED Code \\ \hline
				Output \& Income & RPI, INDPRO, CUMFNS  \\
				Sales & CMRMTSPLx,  RETAILx \\
				Housing & HOUST \\
				Employment &  UNRATE, PAYEMS, USFIRE, CLAIMSx \\
				Prices & CPIAUCSL, CPIULFSL, PCEPI, WPSFD49207 \\ & OILPRICEx \\
				Money &  M2SL\\
				Interest rates & FEDFUNDS \\
				Stock prices & S\&P 500\\
				\hline
			\end{tabular}
	\end{minipage}}
	\label{tab:categories monthly}
\end{table}

\begin{table}[H]
	\centering
	\caption{Medium $(K=55)$ MF-VAR: Linkages between macroeconomic group. Entry $(i,j)$ indicates the number of edges from group $j$ to group $i$.}
	\resizebox{\textwidth}{!}{
		\begin{tabular}{c|r r r r r r r r r |r }
			\hline
			To / From &GDP	& Output	& Sales	& Housing &	Employ- &	Prices	&Money&	Interest &	Stock &	\textit{In-} \\
			& &  \& Income  & & & ment& && Rate & Prices & \textit{degree}\\
			\hline
			GDP & 0 & 3 & 2 & 1 & 4 & 2 & 2 & 0 & 1 & 15\\ 
			Output \& Income & 5 & 36 & 18 & 2 & 30 & 27 & 8 & 9 & 15 & 150 \\ 
			Sales & 4 & 19 & 11 & 5 & 30 & 28 & 6 & 6 & 6 & 115 \\ 
			Housing & 1 & 1 & 2 & 8 & 6 & 0 & 1 & 0 & 0 & 19 \\ 
			Employment & 9 & 36 & 21 & 8 & 70 & 29 & 11 & 14 & 10 & 208\\ 
			Prices & 8 & 45 & 24 & 5 & 45 & 72 & 10 & 15 & 15 & 239\\ 
			Money & 1 & 10 & 4 & 1 & 10 & 12 & 6 & 3 & 3 & 50 \\ 
			Interest Rate & 1 & 6 & 8 & 2 & 14 & 9 & 2 & 6 & 5 & 53\\ 
			Stock Prices & 1 & 8 & 4 & 1 & 6 & 13 & 2 & 2 & 3 & 40\\ 
			\hline
			\textit{Out-degree}&   30 & 164 & 94 & 33 & 215  & 192 & 48 & 55 & 58 & 889  \\
			\hline
	\end{tabular}}
	\label{tab:Variable Selection medium MFVAR per category}
\end{table}

\begin{table}[H]
	\centering
	\caption{Large $(K=91)$ MF-VAR: Linkages between macroeconomic group. Entry $(i,j)$ indicates the number of edges from group $j$ to group $i$. Weekly variables are separated.}
	\resizebox{\textwidth}{!}{
		\begin{tabular}{c|r r r r r r | r r r r |r }
			\hline
			To / From	&GDP&	Output &	Sales&	Housing&	Employ-&	Prices&	CLAIMSx&	Money&	Interest&	Stock &	\textit{In-} \\
			& &  \& Income  & & & ment& && &Rate & Prices & \textit{degree}\\
			\hline
			GDP & 0 & 4 & 2 & 1 & 4 & 1 & 1 & 2 & 2 & 2 & 19 \\ 
			Output \& Income & 7 & 35 & 20 & 4 & 23 & 26 & 13 & 18 & 18 & 16 & 180 \\ 
			Sales & 4 & 20 & 11 & 4 & 23 & 24 & 10 & 11 & 12 & 11 & 130 \\ 
			Housing & 1 & 1 & 2 & 8 & 6 & 0 & 0 & 1 & 1 & 1 & 21 \\ 
			Employment & 6 & 31 & 21 & 5 & 52 & 16 & 10 & 11 & 14 & 8 & 174 \\ 
			Prices & 8 & 47 & 26 & 5 & 28 & 71 & 24 & 17 & 24 & 27 & 277 \\ 
			CLAIMSx & 2 & 16 & 9 & 4 & 9 & 14 & 9 & 8 & 10 & 5 & 86 \\ 
			Money & 2 & 18 & 11 & 4 & 16 & 25 & 7 & 11 & 10 & 8 & 112 \\ 
			Interest Rate & 2 & 20 & 9 & 3 & 16 & 13 & 7 & 9 & 13 & 7 & 99 \\ 
			Stock Prices & 2 & 15 & 11 & 6 & 13 & 25 & 5 & 9 & 7 & 8 & 101 \\ 
			\hline
			\textit{Out-degree}& 34 &  207 & 122 & 44 & 190 & 215 & 86 & 97 & 111 & 93 & 1199 \\
			\hline
	\end{tabular}}
	\label{tab:Variable Selection large MFVAR per category}
\end{table}

\newpage
\subsection{Additional Figures}\label{Appendix Application Figures}

\begin{figure}[h]
	\centering
	\makebox[\textwidth][c]{\includegraphics[scale = 0.55]{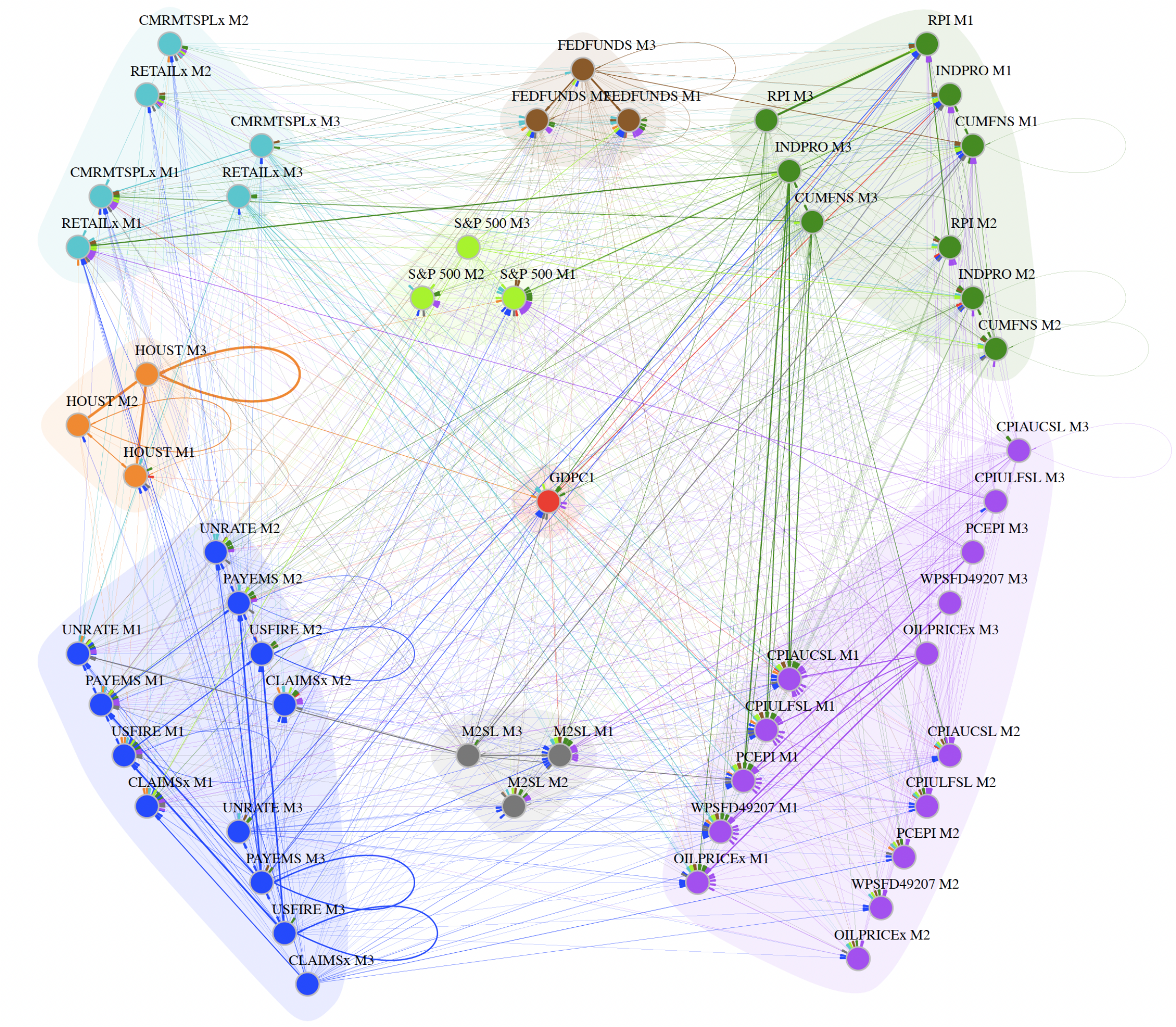}}
	\caption{Medium $(K=55)$ MF-VAR: Directed network: 
		the vertices represent the variables, the edges the nonzero coefficients. The edges' width are proportional to the absolute value of the estimates. Coloring of the vertices and their outgoing edges indicate the macroeconomic categories.}
	
	\label{fig:Medium MFVAR Network}
	\vspace{-50mm}
\end{figure}

\begin{figure}
	\centering
	\makebox[\textwidth][c]{\includegraphics[scale = 0.55]{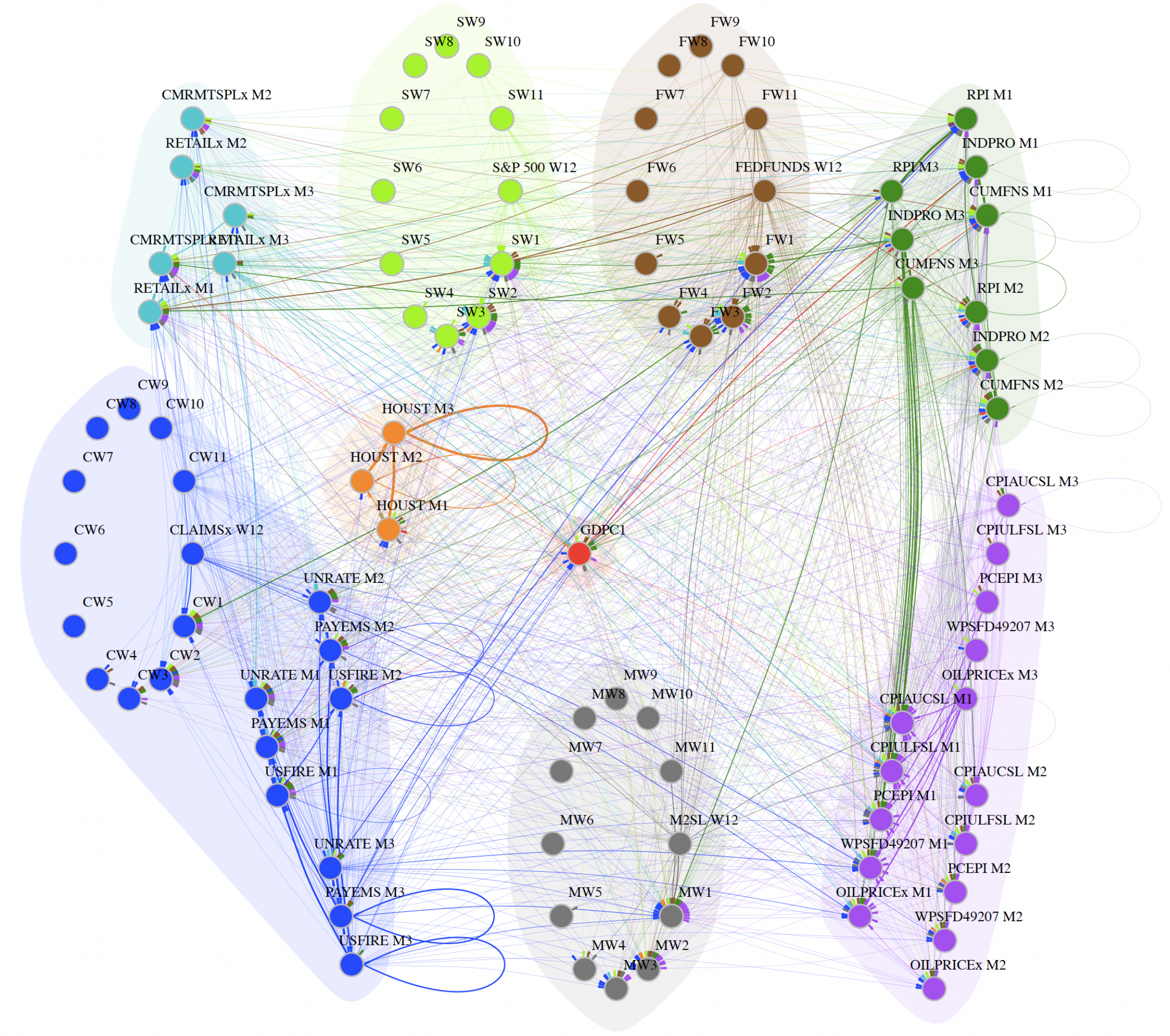}}
	\caption{Large $(K=91)$ MF-VAR: Directed network: the vertices represent the variables, the edges the nonzero coefficients. The edges' width are proportional to the absolute value of the estimates. Coloring of the vertices and their outgoing edges indicate the macroeconomic categories.}
	\label{fig:Large MFVAR Network}
\end{figure}

\begin{figure}
	\begin{subfigure}[]{0.49\textwidth}
		\includegraphics[scale = 0.5]{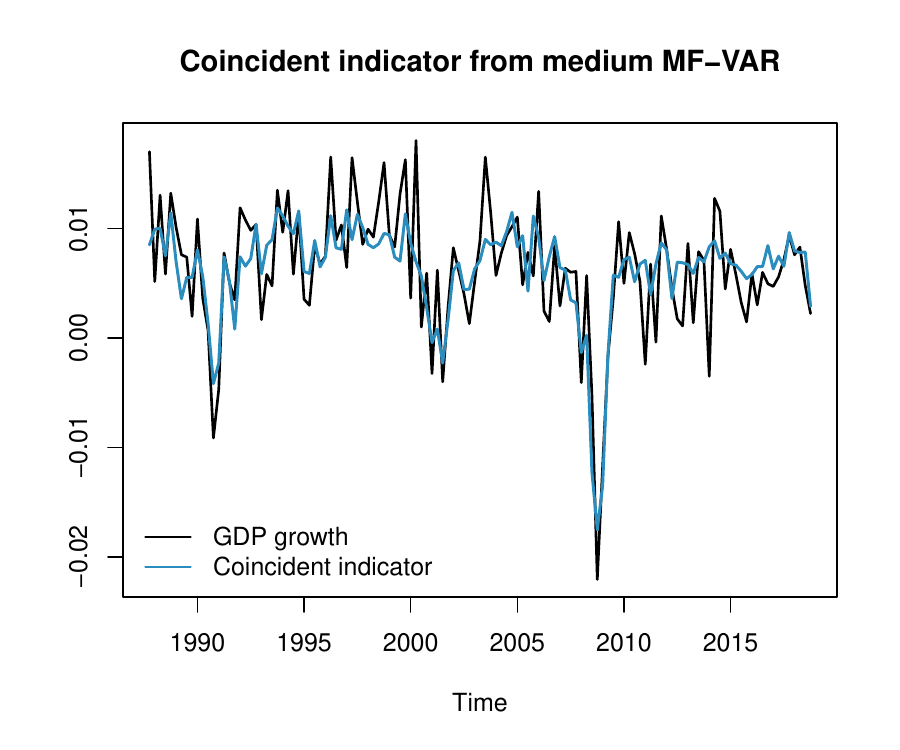}
		\caption{}
	\end{subfigure} 
	\begin{subtable}[]{0.49\linewidth}
		\centering
		\resizebox{0.85\textwidth}{!}{\begin{minipage}{\textwidth}
				\begin{tabular}{l|c c c }
					\hline 
					Variable & \multicolumn{3}{c}{Nowcast} \\ \hline
					RPI & M1 & M2 & M3 \\
					CMRMTSPLx & M1 & M2 & M3 \\
					RETAILx & M1 & M2 & M3 	\\
					HOUST & M1 & M2 & M3 \\
					INDPRO & M1 & M2 & M3 	\\
					CUMFNS & M1 & M2 & M3  	\\
					UNRATE & M1 & M2 & M3 	\\
					PAYEMS &  & M2 & M3 \\	
					USFIRE & & M2 & M3  \\
					CLAIMSx	& M1 & M2 & M3 \\
					CPIAUCSL &  & M2 & 	\\
					CPIULFSL & & M2 & M3 	\\
					PCEPI &  & M2 &  	\\
					WPSFD49207 & M1 & M2 &	 \\
					OILPRICEx & & M2 & 	\\
					M2SL & M1 &  & M3 \\
					FEDFUNDS & & M2 & M3	\\
					S\&P 500 & M1 & M2 & M3 \\
					\hline
				\end{tabular}
		\end{minipage}}
		\caption{} 
	\end{subtable}
	
	\caption{Medium ($K=55$) MF-VAR: Panel (a): U.S.\ GDP growth versus coincident indicator. Panel (b): Selected high-frequency variables for the coincident indicator.}
	\label{fig:GDP vs indicator K = 55}
\end{figure}

\begin{figure}
	\begin{subfigure}[]{0.4\textwidth}
		\includegraphics[scale = 0.5]{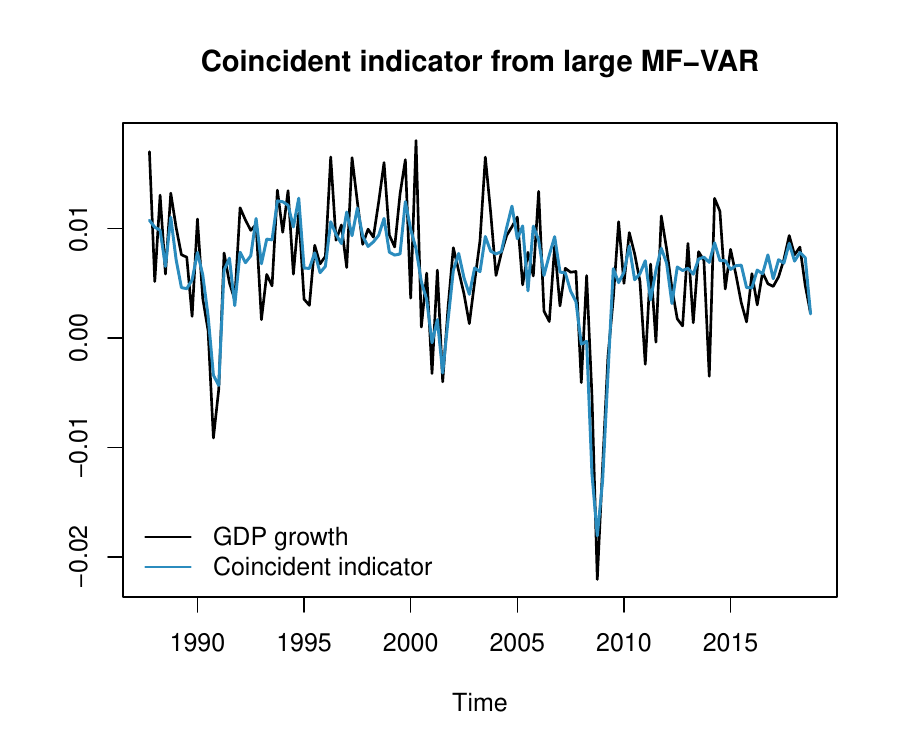}
		\caption{}
	\end{subfigure} 
	\begin{subtable}[]{0.5\textwidth}
		\centering
		\resizebox{0.7\textwidth}{!}{\begin{minipage}{\textwidth}
				\begin{tabular}{l|c c c }
					\hline
					Variable & \multicolumn{3}{c}{Nowcast} \\ \hline
					RPI & M1 & M2 & M3\\
					CMRMTSPLx & M1 & M2 & M3\\
					RETAILx & M1 & M2 & M3	\\
					HOUST & M1 & M2 & M3 \\
					INDPRO & M1 & M2 & M3 \\
					CUMFNS & M1 & M2 & M3	\\
					UNRATE & M1 & M2 & M3 \\
					PAYEMS &  & M2 & M3 \\	
					USFIRE & & & M3 \\
					CPIAUCSL &  & M2 & 	\\
					CPIULFSL & & M2 & M3	\\
					PCEPI &  & M2 & 	\\
					WPSFD49207 & M1 & M2 &	\\
					OILPRICEx 	&M1& M2& \\
					\hline
					CLAIMSx	& W1,W2,W3& W6, W7& W10\\
					M2SL & W2, W4& W5, W6, W7& W10,  W11	\\
					FEDFUNDS & W3, W4&  W5, W7, W8 & W9, W11 \\
					S\&P 500 &  W2, W4&W6, W8,& W9, W10, W11, W12\\
					\hline
				\end{tabular}
		\end{minipage}}
		\caption{}
	\end{subtable}
	
	\caption{Large ($K=91$) MF-VAR: Panel (a): U.S.\ GDP growth versus coincident indicator. Panel (b): Selected high-frequency variables for the coincident indicator.}
	\label{fig:GDP vs indicator K = 91}
\end{figure}

\clearpage

\subsection{Sensitivity analysis: Hierarchical estimator with equal weights}\label{appendix:SA equal weights}
\begin{table}[h]
	\centering
	\caption{Small $(K=22)$ MF-VAR with equal weights: Linkages between macroeconomic group. Entry $(i,j)$ indicates the number of edges from group $j$ to group $i$.}
	\resizebox{0.9\textwidth}{!}{\begin{minipage}{\textwidth}
			\centering
			\begin{tabular}{c|r r r r r |r }
				\hline
				To/From & GDP & Output \& Income & Housing	& Employment	& Prices &	 \textit{In-degree}\\
				\hline
				GDP & 0 & 4 & 1 & 5 & 0 & 10  \\ 
				Output \& Income & 4 & 32 & 6 & 24 & 16 & 82 \\ 
				Housing & 1 & 2 & 9 & 9 & 0 & 21\\ 
				Employment & 5 & 37 & 6 & 68 & 11 & 127 \\ 
				Prices & 3 & 16 & 1 & 17 & 9 & 46 \\ 
				
				\hline
				\textit{Out-degree}	&13 & 91 & 23 & 123 & 36 & 286 \\
				\hline
			\end{tabular}
	\end{minipage} }
	\label{tab:Variable Selection small MFVAR per category equal weights}
\end{table}

\begin{table}[h]
	\centering
	\caption{Medium $(K=55)$ MF-VAR with equal weights: Linkages between macroeconomic group. Entry $(i,j)$ indicates the number of edges from group $j$ to group $i$.}
	\resizebox{\textwidth}{!}{
		\begin{tabular}{c|r r r r r r r r r |r }
			\hline
			To / From &GDP	& Output	& Sales	& Housing &	Employ- &	Prices	&Money&	Interest &	Stock &	\textit{In-} \\
			& &  \& Income  & & & ment& && Rate & Prices & \textit{degree}\\
			\hline
			GDP & 0 & 3 & 3 & 1 & 5 & 1 & 2 & 1 & 2 & 18 \\ 
			Output \& Income & 4 & 68 & 31 & 7 & 60 & 55 & 17 & 20 & 19 & 281\\ 
			Sales & 4 & 36 & 24 & 7 & 59 & 55 & 14 & 15 & 16 & 230 \\ 
			Housing & 1 & 1 & 2 & 9 & 8 & 0 & 1 & 0 & 0 & 22\\ 
			Employment & 5 & 48 & 44 & 10 & 107 & 53 & 23 & 22 & 21 & 333 \\ 
			Prices & 10 & 101 & 55 & 6 & 87 & 155 & 17 & 30 & 31 & 492\\ 
			Money & 1 & 17 & 7 & 3 & 22 & 18 & 8 & 6 & 3 & 85 \\ 
			Interest Rate & 0 & 3 & 11 & 1 & 16 & 21 & 1 & 9 & 7 & 69\\ 
			Stock Prices & 0 & 19 & 8 & 5 & 11 & 26 & 8 & 9 & 9 & 95 \\ 
			\hline
			\textit{Out-degree}& 25 & 296 & 185 & 49 & 375 & 384 & 91 & 112 & 108 & 1625 \\
			\hline
	\end{tabular}}
	\label{tab:Variable Selection medium MFVAR per category equal weights}
\end{table}

\begin{table}[H]
	\centering
	\caption{Large $(K=91)$ MF-VAR with equal weights: Linkages between macroeconomic group. Entry $(i,j)$ indicates the number of edges from group $j$ to group $i$. Weekly variables are separated.
	}
	\resizebox{\textwidth}{!}{
		\begin{tabular}{c|r r r r r r | r r r r |r }
			\hline
			To / From	&GDP&	Output &	Sales&	Housing&	Employ-&	Prices&	CLAIMSx&	Money&	Interest&	Stock &	\textit{In-} \\
			& &  \& Income  & & & ment& && &Rate & Prices & \textit{degree}\\
			\hline
			GDP & 0 & 4 & 4 & 1 & 5 & 1 & 2 & 5 & 5 & 3 & 30 \\ 
			Output \& Income & 6 & 68 & 35 & 7 & 45 & 56 & 49 & 62 & 52 & 57 & 437 \\ 
			Sales & 5 & 32 & 24 & 8 & 42 & 51 & 32 & 37 & 38 & 39 & 308\\ 
			Housing & 1 & 1 & 2 & 9 & 8 & 0 & 0 & 0 & 2 & 1 & 24 \\ 
			Employment & 4 & 35 & 33 & 4 & 69 & 26 & 35 & 27 & 38 & 28 & 299 \\ 
			Prices & 10 & 101 & 50 & 22 & 57 & 152 & 93 & 72 & 78 & 104 & 739 \\ 
			CLAIMSx & 2 & 60 & 41 & 13 & 35 & 49 & 123 & 110 & 116 & 87 & 636 \\ 
			Money & 2 & 63 & 29 & 22 & 45 & 71 & 103 & 140 & 123 & 94 & 692 \\ 
			Interest Rate & 3 & 74 & 28 & 3 & 52 & 54 & 84 & 115 & 135 & 105 & 653 \\ 
			Stock Prices & 3 & 52 & 33 & 18 & 48 & 79 & 59 & 111 & 106 & 121 & 630\\ 
			\hline
			\textit{Out-degree}&  36 & 490 & 279 & 107 & 406  & 539 & 580 & 679 & 693 & 639 & 4448 \\
			\hline
	\end{tabular}}
	\label{tab:Variable Selection large MFVAR per category equal weights}
\end{table}

\begin{table}[h]
	\centering
	\caption{Correlation between U.S.\ GDP growth and the coincident indicators for the small $(K= 22)$,  medium $(K= 55)$ and large $(K= 91)$ MF-VAR groups with equal weights. }
	\resizebox{0.9\textwidth}{!}{\begin{minipage}{\textwidth}
			\centering
			\begin{tabular}{l c c c}
				\hline
				Type of coincident indicator& $K=22$ & $K=55$ & $K=91$ \\
				\hline
				Nowcasting relations M1& 0.6150 & 0.2658 &  0.5528\\
				Nowcasting relations M1 + M2 & 0.7288 & 0.7780 & 0.7769 \\
				All nowcasting relations & 0.7429 & 0.7705 &  0.7956\\
				All variables & 0.7216 & 0.7564 & 0.7557 \\
				\hline
			\end{tabular}
			\label{tab:correlations overview equal weights}
	\end{minipage}}
\end{table}

\subsection{Sensitivity analysis: Daily data}\label{Appendix:SA HF data}

\begin{table}[H]
	\centering
	\caption{Ultra-large $(K=187)$ MF-VAR: Linkages between macroeconomic group. Entry $(i,j)$ indicates the number of edges from group $j$ to group $i$. Weekly and daily variables are separated.}
	\resizebox{\textwidth}{!}{
		\begin{tabular}{c|r r r r r r | r r | r r |r }
			\hline
			To / From	&GDP&	Output &	Sales&	Housing&	Employ-&	Prices&	CLAIMSx&	Money&	Interest&	Stock &	\textit{In-} \\
			& &  \& Income  & & & ment& && &Rate & Prices & \textit{degree}\\
			\hline
			GDP & 0 & 5 & 3 & 1 & 4 & 2 & 2 & 2 & 2 & 3 & 24\\ 
			Output \& Income & 6 & 42 & 23 & 5 & 27 & 30 & 14 & 18 & 26 & 29 & 220 \\ 
			Sales & 5 & 21 & 16 & 4 & 24 & 26 & 11 & 12 & 17 & 18 & 154\\ 
			Housing & 1 & 0 & 2 & 8 & 6 & 0 & 0 & 0 & 2 & 6 & 25\\ 
			Employment & 7 & 31 & 22 & 5 & 53 & 19 & 10 & 8 & 22 & 25 & 202\\ 
			Prices & 7 & 58 & 29 & 8 & 34 & 89 & 25 & 20 & 38 & 44 & 352 \\ 
			CLAIMSx & 3 & 16 & 10 & 5 & 12 & 21 & 9 & 8 & 16 & 14 & 114 \\ 
			Money & 2 & 19 & 12 & 4 & 18 & 24 & 9 & 14 & 14 & 15 & 131\\ 
			Interest Rate & 3 & 25 & 18 & 6 & 24 & 29 & 13 & 17 & 40 & 22 & 197\\ 
			Stock Prices & 5 & 29 & 19 & 9 & 30 & 43 & 10 & 16 & 27 & 38 &226 \\ 
			\hline
			\textit{Out-degree}&   39 & 246 & 154 & 55 & 232 & 283 & 103 & 115& 204 & 214 &  1645\\
			\hline
	\end{tabular}}
	\label{tab:Variable Selection mega MFVAR per category}
\end{table}

\begin{table}[h]
	\centering
	\caption{Correlation between U.S.\ GDP growth and the coincident indicators for the large $(K= 91)$ and ultra-large ($K=187$) MF-VAR. }
	\resizebox{0.9\textwidth}{!}{\begin{minipage}{\textwidth}
			\centering
			\begin{tabular}{l c c}
				\hline
				Type of coincident indicator & $K=91$ & $K=187$\\
				\hline
				Nowcasting relations M1&  0.4887 & 0.6617\\
				Nowcasting relations M1 + M2 & 0.6680 & 0.7303\\
				All nowcasting relations & 0.7928 & 0.7893 \\
				All variables & 0.7557 & 0.7337 \\
				\hline
			\end{tabular}
			\label{tab:correlations overview large+mega}
	\end{minipage}}
\end{table}

\subsection{Sensitivity analysis: Forecast Comparison}\label{Appendix:SA Forecast Comparison}
We use a rolling-window set-up with window size $T_1 = 105$, providing us 20 quarterly observations for forecast comparison. 
For each rolling window, we select the value of $\lambda_\beta$ that minimizes the one-step-ahead squared forecast error for our main variable of interest \textit{GDPC1}.
The MSFE for \textit{GDPC1} for the unrestricted hierarchical estimator across all MF-VARs is given in Table \ref{tab:MSFE GDP}.\footnote{We have also performed the forecasting exercise with the restricted hierarchical estimator imposing the nowcasting restrictions (through GLS). In line with our simulation study, the GLS version did not result in an improved point forecast and thus is omitted.}

\begin{table}[H]
	\caption{Rolling out-of-sample one-step-ahead MSFE of \textit{GDPC1}. Forecast methods in the 75\% Model Confidence Set (MCS) are in bold. Standard errors are in parentheses.}
	\centering
	\resizebox{0.9\textwidth}{!}{\begin{minipage}{\textwidth}
			\centering
			\begin{tabular}{l c c c c }
				\hline
				Estimator & {Univariate} & \multicolumn{3}{c}{Multivariate} \\
				&& $K = 22$ & $K=55$ & $K=91$ \\
				\hline
				AR &  0.4137 (0.1880)\\
				RW & 0.6969 (0.3636) & \\
				\\
				Quarterly VAR & & 0.3070 (0.0907)& 0.3624 (0.1094) \\
				OLS   & &0.5884 (0.1977) & 0.9285 (0.2572) &  2.6267 (0.9511) \\
				Ridge & &0.2307 (0.0977)& 0.2191 (0.0761)& \textbf{0.1630} (0.0491) \\
				Lasso & &\textbf{0.1674} (0.0886) & 0.1565 (0.0773) & \textbf{0.1326} (0.0666)  \\
				Hierarchical & &0.1678 (0.0824) & \textbf{0.1477} (0.0691)& \textbf{0.1062} (0.0379) \\
				\hline
			\end{tabular}
			\label{tab:MSFE GDP}
	\end{minipage}}
\end{table}

\end{document}